\documentclass{article}

\usepackage{graphicx}
\usepackage{epsfig}
\usepackage{amssymb}
\usepackage{amsmath}
\usepackage{amsfonts}
\usepackage{color} 
\usepackage{cite}
 
\begin{document}
\begin{center}

{\LARGE \bf Quasinormal modes 
of hot, cold and bald
\\[1ex] 
 Einstein-Maxwell-scalar black holes}
\vspace{0.8cm}
\\
{{\bf Jose Luis Bl\'azquez-Salcedo$^{\Diamond}$,
Carlos A. R. Herdeiro$^{\ddagger}$,  \\
Sarah Kahlen$^{\Diamond}$, Jutta Kunz$^{\Diamond}$,
Alexandre M. Pombo$^{\ddagger}$ and 
Eugen Radu$^{\ddagger}$ 
}
\vspace{0.3cm}
\\
$^{\Diamond}${\small Institut f\"ur  Physik, Universit\"at Oldenburg, Postfach 2503,
D-26111 Oldenburg, Germany}
\vspace{0.3cm}
\\
$^{\ddagger }${\small Departamento de Matem\'atica da Universidade de Aveiro and } \\ {\small  Centre for Research and Development  in Mathematics and Applications (CIDMA),} \\ {\small    Campus de Santiago, 3810-183 Aveiro, Portugal}
}
\vspace{0.3cm}
\end{center}

\date{\today}

\begin{abstract}
{Einstein-Maxwell-scalar models allow for different classes of black hole solutions, depending on the non-minimal coupling function $f(\phi)$ employed, between the scalar field  and the  Maxwell invariant.}
Here, we address the linear mode stability of the black hole solutions obtained
recently for a quartic coupling function, $f(\phi)=1+\alpha\phi^4$ \cite{Blazquez-Salcedo:2020nhs}.
Besides the bald Reissner-Nordstr\"om solutions, this coupling allows for two branches of
scalarized black holes, termed cold and hot, respectively.
For these three branches of black holes we calculate the spectrum of quasinormal modes.
It consists of polar scalar-led modes, polar and axial electromagnetic-led modes,
and polar and axial gravitational-led modes. We demonstrate that the only unstable mode
present is the radial scalar-led mode of the cold branch. 
Consequently, the bald Reissner-Nordstr\"om branch and the hot scalarized branch are both mode-stable.
The non-trivial scalar field in the scalarized background solutions leads to
the breaking of the degeneracy between axial and polar modes present for Reissner-Nordstr\"om solutions.
This isospectrality is only slightly broken on the cold branch, but it is strongly broken on the hot branch.
\end{abstract}


\tableofcontents

\newpage

\section{Introduction}

The phenomenon of spontaneous scalarization of black holes has received much interest in recent years.
{In certain scalar-tensor theories,} 
the well-known black holes of General Relativity {(GR)} remain solutions of the
field equations, while in certain regions of the parameter space, additional branches of black hole solutions arise,
that are endowed with scalar hair.
Spontaneous scalarization can, for instance, be charge-induced,
when a scalar field is suitably coupled to the Maxwell invariant $F^2$
\cite{Herdeiro:2018wub}.
{Whereas the onset of the instability is universal, the}
 properties of the resulting scalarized black holes and the branch structure of the solutions then depend 
significantly on the coupling function $f(\phi)$
\cite{Blazquez-Salcedo:2020nhs,Herdeiro:2018wub,Myung:2018vug,Boskovic:2018lkj,Myung:2018jvi,Fernandes:2019rez,Brihaye:2019kvj,Herdeiro:2019oqp,Myung:2019oua,Astefanesei:2019pfq,Konoplya:2019goy,Fernandes:2019kmh,Herdeiro:2019tmb,Zou:2019bpt,Brihaye:2019gla}. 

{Einstein-Maxwell-scalar (EMs) models include two distinctive classes, depending on the choice of $f(\phi)$: models that have black holes with scalar hair only (and do not allow the GR solutions) or models that allow both  the GR solutions and new hairy black holes. The latter splits into  two further sub-classes: models wherein the GR black  holes are unstable, in  some region of parameter space, against the tachyonic instability  that promotes scalarization and models wherein the GR black holes are never afflicted by this instability. The last two sub-classes  have been labelled as classes IIA and IIB, respectively,  in~\cite{Astefanesei:2019pfq}.}

{Many previous studies on EMs BHs in the last few years have been motivated by charge-induced spontaneous scalarization, and thus have focused on black holes of class IIA 
\cite{Herdeiro:2018wub,Myung:2018vug,Boskovic:2018lkj,Myung:2018jvi,Fernandes:2019rez,Brihaye:2019kvj,Herdeiro:2019oqp,Myung:2019oua,Astefanesei:2019pfq,Konoplya:2019goy,Fernandes:2019kmh,Herdeiro:2019tmb,Zou:2019bpt,Brihaye:2019gla}. }
In the static case, the branch of Reissner-Nordstr\"om (RN) black holes then develops 
a zero mode at a critical charge to mass ratio $q_{\text{cr}}(\alpha)$,
that depends on the coupling strength $\alpha$. 
At this bifurcation point, a new branch of scalarized black holes emerges. 
The zero mode of the RN black hole branch 
then turns into an unstable mode, while the new scalarized branch may or may not be stable,
depending on the coupling function
\cite{Myung:2018vug,Myung:2018jvi,Myung:2019oua,Zou:2019bpt}.
Of course, unstable excited scalarized branches exist as well.
Let us note at this point that this observed pattern of branches and instabilities
is very similar to the one seen in curvature-induced spontaneously scalarized black holes
\cite{Doneva:2017bvd,Antoniou:2017acq,Silva:2017uqg,Antoniou:2017hxj,Blazquez-Salcedo:2018jnn,Doneva:2018rou,Minamitsuji:2018xde,Silva:2018qhn,Brihaye:2018grv,Doneva:2019vuh,Myung:2019wvb,Cunha:2019dwb,Macedo:2019sem,Hod:2019pmb,Collodel:2019kkx,Bakopoulos:2020dfg,Blazquez-Salcedo:2020rhf,Blazquez-Salcedo:2020caw}. 

Recently, we have considered {an  EMs model} 
employing the quartic 
coupling function $f(\phi)=1+\alpha \phi^4$, which qualifies as class IIB,
since the RN branch does not become unstable against scalar perturbations
\cite{Blazquez-Salcedo:2020nhs}.
Instead, we have observed the following interesting pattern:
Close to the extremal RN solution, corresponding to the mass to charge ratio $q=1$,
a first branch of scalarized black holes emerges.
This first branch then exists in the range $q_{\text{min}}(\alpha) \le q \le 1$.
At $q_{\text{min}}(\alpha)$, it bifurcates with a second branch of scalarized black holes,
which extends throughout the interval $q_{\text{min}}(\alpha) \le q \le q_{\text{max}}(\alpha)$,
and ends in an extremal singular solution at
$q_{\text{max}}(\alpha) > 1$.
Considering the properties of these three branches, we have termed the RN branch
{as the \textit{bald} branch, since it does not carry scalar hair, while we have termed the first
and the second branches as the \textit{cold} and the \textit{hot} branches,} respectively, 
according to the black hole horizon temperatures.

In this first study \cite{Blazquez-Salcedo:2020nhs}, 
we have also addressed the radial modes of the black holes for these three branches,
since the radial modes signal instabilities with respect to scalar perturbations
and thus the onset of scalar hair.
While our analysis has shown that there are no unstable radial modes on the RN branch 
and on the hot scalarized branch, we have found that
the cold scalarized branch develops an unstable mode close to the point $q=1$. This instability is present throughout the interval $q_{\text{min}}(\alpha) < q < 1$ and ends with a zero mode
at the bifurcation point $q_{\text{min}}(\alpha)$ with the hot scalarized branch.
Thus, the cold scalarized branch is clearly unstable. However, {it remained an open issue}
whether there are really two stable coexisting branches, the bald RN branch and the hot scalarized branch.
This question has motivated our present investigation.

Here, we study linear mode stability of the black holes on all three branches.
To that end, we calculate the lowest quasinormal modes for each type of mode.
In particular, since the theory has scalar and vector fields coupled to gravity,
we have to consider the perturbations of all these fields.
In the presence of non-trivial background fields, 
i.e., when the black hole solutions carry scalar hair and electromagnetic charge,
the different types of perturbations generically couple to each other,
leading to scalar-led, electromagnetic-led and gravitational-led modes
instead of pure scalar, electromagnetic or gravitational modes,
which one would find for a Schwarzschild black hole, for instance.
Since parity even (polar) and parity odd (axial) perturbations decouple,
we arrive at the following set of modes when expanding in spherical harmonics:
polar scalar-led $l \ge 0$ modes, axial and polar electromagnetic-led $l \ge 1$ modes,
and axial and polar gravitational-led $l \ge 2$ modes.

In Section II, we present the {EMs} theory studied
and the general set of equations. We specify the Ansatz for the spherically symmetric
background solutions and give the resulting set of ordinary differential equations (ODEs).
We discuss the asymptotic expansions for asymptotically flat black hole solutions 
with a regular horizon in Section  III. Here, we also recall basic properties of the three branches of
black hole solutions. In Section IV, we formulate the sets of perturbation equations
for spherical, axial and polar perturbations, deferring more details to the Appendix.
We present our numerical results for the quasinormal modes in Section V.
Here, we show that no further unstable modes arise on any of the three branches.
Moreover, we discuss the breaking of isospectrality,
i.e., the splitting of the axial and polar electromagnetic-led and gravitational-led modes,
caused by the presence of the scalar hair. We conclude in Section V.

\section{EMs theory}

We consider EMs theory described by the action
\begin{equation}\label{E21}
	 \mathcal{S}=\int d^4 x \sqrt{-g} 
	 \Big[R-2\partial _\mu \phi \partial ^\mu \phi -f (\phi) F_{\mu \nu} F^{\mu \nu} \Big]\ ,
\end{equation}
where $R$ is the Ricci scalar, $\phi$ is a real scalar field, and $F_{\mu\nu}$ is the Maxwell field strength tensor.
The coupling between the scalar and Maxwell fields is determined by the coupling function $f(\phi)$,
for which we assume a quartic dependence,
\begin{equation}
f(\phi) = 1 +\alpha \phi^4 \ .
\label{coup}
\end{equation}
For a positive coupling constant $\alpha$,
the global minimum of the coupling function is at $\phi=0$.

The set of coupled field equations is obtained via the variational principle and reads
\begin{eqnarray}
\label{Eins} 
R_{\mu\nu}-\frac{1}{2}g_{\mu\nu}R=
T_{\mu\nu}^{\phi}+T_{\mu\nu}^{EM} \ , \\
\nabla_{\mu}(\sqrt{-g} f(\phi) F^{\mu\nu} )
=0 \ , \label{Mxw}
\\
\frac{1}{\sqrt{-g}}\partial_{\mu}
 (\sqrt{-g}g^{\mu\nu}\partial_{\nu}\phi )
 = \dot f(\phi) F_{\mu\nu}F^{\mu\nu} \ ,
\label{Klein}
\end{eqnarray}
where $\dot f(\phi) = d  f(\phi) /d \phi$, and
$T_{\mu\nu}^{\phi}$ and $T_{\mu\nu}^{EM}$
are the scalar stress-energy tensor
and the electromagnetic stress-energy tensor, respectively.
\begin{eqnarray}
T_{\mu\nu}^{\phi}
\equiv\frac{1}{2}\partial_{\mu}\phi\partial_{\nu}\phi
-\frac{1}{2}g_{\mu\nu}
 \left( \frac{1}{2}(\partial_\alpha \phi)^2+V(\phi)  \right) \ , \\
T_{\mu\nu}^{EM}\equiv2 f(\phi)
\left(F_{\mu\alpha}F_{\nu}^{\,\,\alpha}
-\frac{1}{4}g_{\mu\nu}F^2\right)\ .
\end{eqnarray}

To construct static spherically symmetric EMs black holes, 
we employ the line element 
\begin{eqnarray}
\label{stab1}
ds^2=-g(r) dt^2 + \frac{dr^2}{1-2m(r)/r} +r^2(d\theta^2+\sin^2 \theta d\varphi^2) \ ,
\end{eqnarray}
where the metric functions $g$ and $m$ depend on the radial coordinate $r$.
The black holes are supposed to carry electric charge and, in the scalarized case, also scalar charge.
We therefore parametrize the gauge potential and the scalar field by
\begin{eqnarray}
\label{stab2}
A&=&  a_0(r) dt \ , \nonumber \\ \phi&=&\phi_0(r) \ , 
\end{eqnarray} 
where $a_0$ and $\phi_0$ are the electric and the scalar field function,
respectively, which depend on the radial coordinate $r$.
Inserting this Ansatz into the general set of EMs equations (\ref{Eins})-(\ref{Klein}),
we obtain the following set of ODEs
for the functions $g$, $m$, $a_{0}$ and $\phi_{0}$: 
\begin{eqnarray}
\label{stateq}
\partial_r\delta &=&-\frac{r}{4}\, \left( \partial_r\phi_{0}  \right)^{2} \ , \nonumber \\ 
\partial_r m &=&\frac{1}{2}\, f(\phi_{0}) {e}^{2 \delta} {r}^{2} \left( \partial_ra_0  \right)^{2}+\frac{r}{8} \left(r-2m\right)  \left(\partial_r\phi_{0}\right)^{2} \ , \nonumber \\
\partial^2_r\phi_{0} &=& \frac{e^{2\delta} r}{r-2m }
\left( r(\partial_r\phi_{0})  f(\phi_{0}) -2\dot  f(\phi_{0}) \right)
\left( \partial_r a_0\right)^{2}  \nonumber \\ 
& & +2 \frac{m-r}{\left(r-2m\right)r} (\partial_r\phi_{0}) \ , \nonumber \\
\partial^2_r a_0 &=& \left(\frac{1}{4}\, \left(\partial_r\phi_{0}\right) ^{2}r
-\frac{\dot f(\phi_0)}{f(\phi_0)} (\partial_r\phi_{0})-\frac{2}{r} \right) (\partial_r a_0) \ , 
\end{eqnarray}
where we have introduced the function $\delta(r)$ via $g=\left(1-\frac{2m}{r}\right)e^{-2\delta}$.
Inserting the first equation into the last yields a
first integral for the electromagnetic field 
\begin{eqnarray}
\partial_r a_0 = \frac{Q}{f(\phi_0) e^{\delta}r^2} \ ,
\end{eqnarray}
where $Q$ is the electric charge of the black holes.
With this first integral, the above set of equations can be simplified.

\section{EMs black holes}

We here briefly recall the properties of static spherically symmetric electrically charged
EMs black hole solutions with the quartic coupling function (\ref{coup}).
First of all, the RN black hole, which is also a solution of the EMs equations, is given by
\begin{eqnarray}
g=1-\frac{2M}{r}+\frac{Q^2}{r^2}  \ , \
m=M-\frac{Q^2}{2r} \ , \
a_0 = -\frac{Q}{r} \ , \
\phi_0 = 0 \ .
\label{RN_sol}
\end{eqnarray} 

The scalarized EMs solutions are obtained numerically \cite{Blazquez-Salcedo:2020nhs}.
Their asymptotic behavior yields their global charges
\begin{eqnarray}
\label{inf_static}
g &=&1-\frac{2M}{r}+ O(r^{-2}) \ , \nonumber \\
m &=&M-\left(Q^2+Q_s^2/4\right)\frac {1}{2r}+O(r^{-2}) \ , \nonumber \\
\phi_{0} &=&\frac{Q_s}{r} + \left(M Q_s - Q^2\frac{\dot f(0)}{f(0)} \right)\frac{1}{r^2} + O(r^{-3}) \ , \nonumber \\
a_0 &=&-\frac{Q}{r} + \frac{ Q Q_s}{2r^2}\frac{\dot f(0)}{f(0)} + O(r^{-3}) \ ,
\end{eqnarray}
with black hole mass $M$, electric charge $Q$, and scalar charge $Q_s$.
Note that there is no conservation law for the scalar field, and that
the existence of a horizon imposes a non-trivial relation $Q_s = Q_s(M,Q)$
\cite{Herdeiro:2015waa}.

Close to the horizon $r=r_H$, the global charges have the expansion
\begin{eqnarray}
\label{hor_static}
g &=&g_1 \left(r-r_H\right) + O(\left(r-r_H\right)^2) \ , \nonumber \\
m &=&\frac{r_H}{2}+\frac{Q^2}{2r_H^2 f(\phi_H)}\left(r-r_H\right) + O(\left(r-r_H\right)^2) \ , \nonumber \\
\phi_0 &=&\phi_H-\frac{2 Q^2 \dot f(\phi_H)}{r_H f(\phi_H) (r_H^2 f(\phi_H) -Q^2)}\left(r-r_H\right) +O(\left(r-r_H\right)^2) \ , \nonumber  \\
a_0 &=&-\Psi_H+Q\sqrt{\frac{g_1}{r_H f(\phi_H) (r_H^2 f(\phi_H) -Q^2)}}\left(r-r_H\right)+O(\left(r-r_H\right)^2) \ ,
\end{eqnarray}
where $\phi_0(r_H)=\phi_H$ and $a_0(\infty)-a_0(r_H)=\Psi_H$
are the value of the scalar field and the electrostatic potential at the horizon.
Further physically relevant horizon properties
are, for instance, the temperature $T_H$ and the horizon area $A_H$, respectively given by
\begin{eqnarray}
T_H&=& \frac{1}{4\pi}\sqrt{\frac{g_1}{r_H^{3}  f(\phi_H) }(r_H^2   -Q^2 f(\phi_H))} \ , \nonumber \\
A_H&=& 4\pi r_H^2 \ .
\end{eqnarray}

\begin{figure}[h!]
			 \centering
	 		 \includegraphics[width=0.35\linewidth,angle=-90]{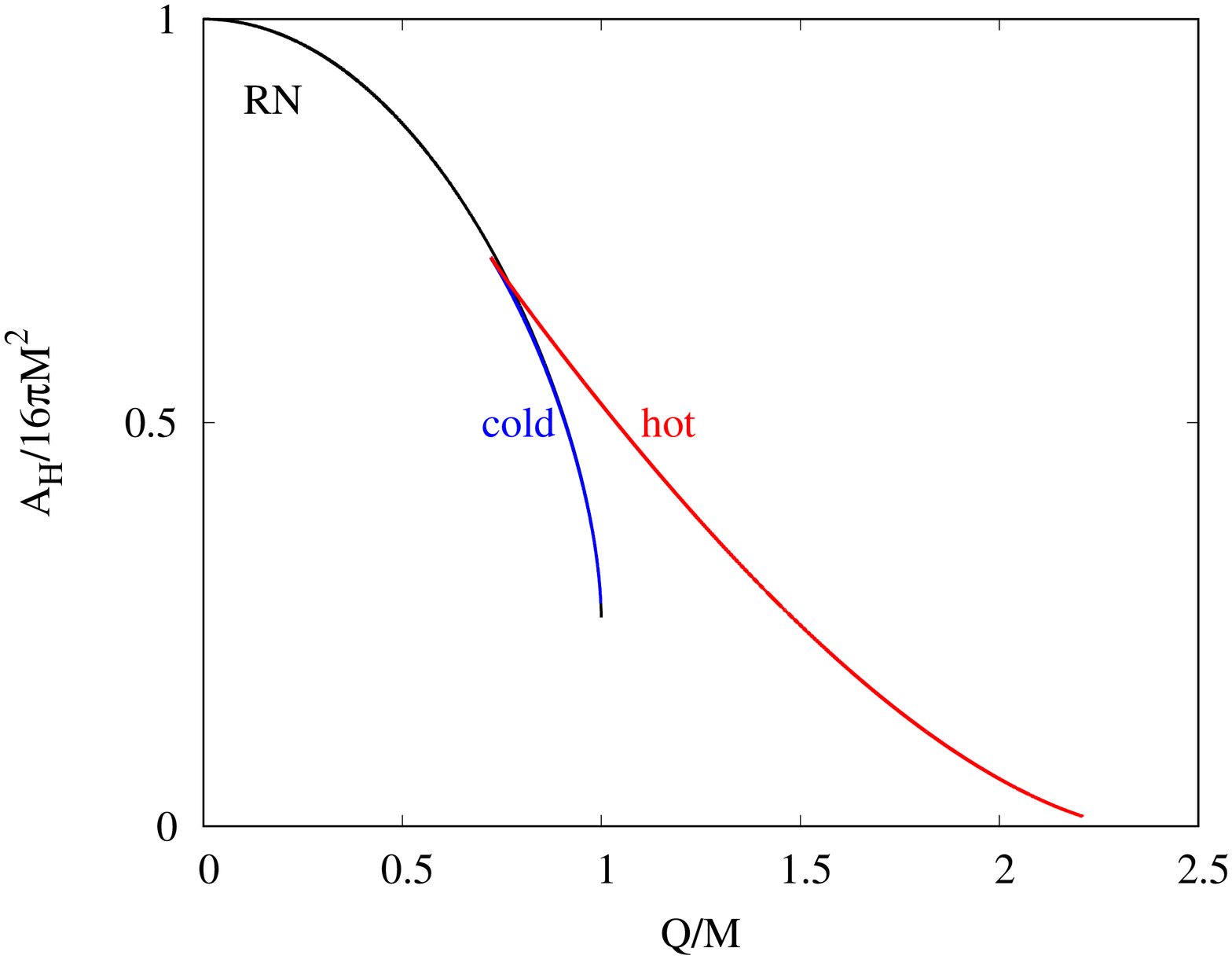}
	 		 \includegraphics[width=0.35\linewidth,angle=-90]{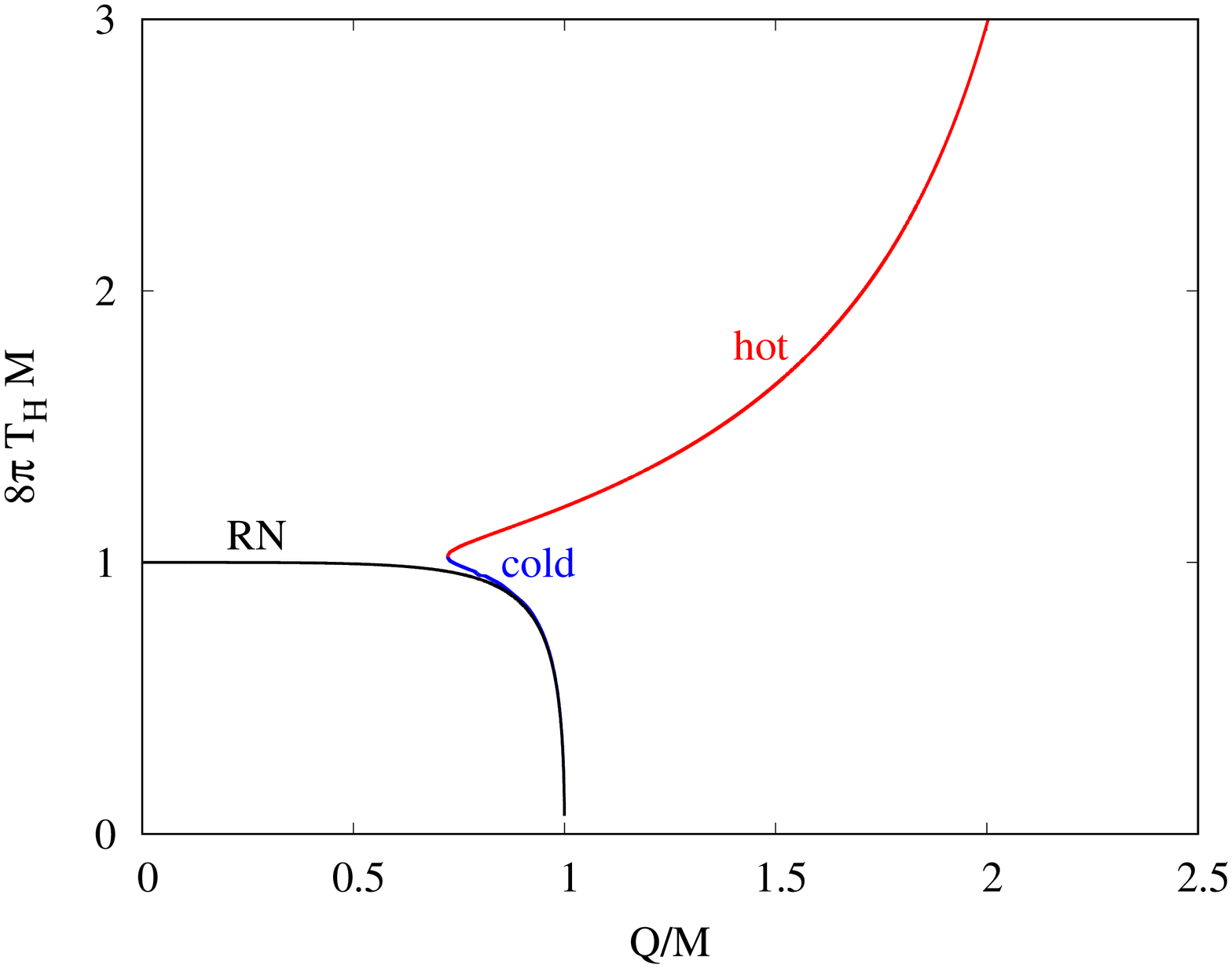}
	 		 \caption{
Scalarised black holes (red, blue) for $\alpha=200$ and RN black holes (black):
(left) reduced area vs. reduced charge; (right) reduced temperature vs. reduced charge. 
Scalarised black holes are cold (blue) on the first branch and hot (red) on the second branch.
}
\label{fig:bh}
\end{figure}

The EMs black hole solutions can be characterized by three dimensionless quantities:
by their charge to mass ratio $q$, their reduced horizon area $a_H$ and their
reduced horizon temperature $t_H$:
\begin{equation}
q\equiv \frac{Q}{M} \ , \qquad  a_H\equiv \frac{A_H}{16\pi M ^2}\ ,\qquad t_H\equiv 8\pi M T_H  \ .
\end{equation}
In the following we will focus on the case with $\alpha=200$, since for other values of the coupling constant the general properties of the black holes are qualitatively similar.
We illustrate the EMs black hole branches in Fig.~\ref{fig:bh},
where we exhibit their reduced area $a_H$ (left) and reduced temperature $t_H$ (right) versus their
charge to mass ratio $q$ \cite{Blazquez-Salcedo:2020nhs}.
The RN black holes are shown in black. The first scalarized branch emerges close to the
extremal RN black hole solution. Along the branch, the mass to charge ratio $q$ decreases,
while the reduced area $a_H$ and temperature $t_H$ increase.
At a minimal value of $q$, the first branch bifurcates with the second branch.
Along the second branch, $a_H$ decreases while $t_H$ increases monotonically
with increasing $q$. Thus, the first branch represents the cold (blue) branch,
and the second branch the hot (red) branch.

\section{Linear perturbation theory}

Previously, we have addressed the \textit{radial} stability of these EMs black holes,
by looking for radially unstable modes \cite{Blazquez-Salcedo:2020nhs}.
Our analysis has revealed radial stability for the RN branch and for the hot scalarized branch.
In contrast, for the cold scalarized branch, we have found an unstable radial mode.
Here, our goal is more ambitious, since we want to {fully clarify} the
 linear mode stability of the
RN and the hot scalarized branch. {As we shall see these are both stable.}

To show linear mode stability, we will evaluate the quasinormal mode spectrum
of the black holes on these branches. For completeness, we will consider the spectrum on all 
three branches, including the unstable cold scalarized branch.
The symmetry of the background solutions suggests to consider
perturbations for the following three cases: 
purely spherical perturbations ($l=0$), axial or odd-parity ($(-1)^{l+1}$) perturbations,
and polar or even-parity ($(-1)^l$) perturbations.

\subsection{Spherical perturbations}

To study the spherical perturbations, we have to perturb all the fields: the scalar field,
the electromagnetic field and the metric. 
For the scalar and the electromagnetic field,
we introduce the perturbation functions $\phi_1$ and $F_{a_0}$,
and employ the Ansatz
\begin{eqnarray}
\label{stab3}
\phi=\phi_0(r) + \epsilon e^{-i \omega t} \phi_1(r) \ , \\
\label{stab3a}
A=   a_0(r) (1+\epsilon e^{-i \omega t} F_{a_0}(r)) dt \ .
\end{eqnarray} 
For the metric, we introduce the perturbations functions $F_t$ and $F_r$
and employ the Ansatz
\begin{eqnarray}
\label{metric_l0_pert}
ds^2=-g(r) (1+\epsilon e^{-i \omega t} F_t(r)) dt^2 + \frac{1+\epsilon e^{-i \omega t} F_r(r)}{1-2m(r)/r}dr^2 +r^2(d\theta^2+\sin^2 \theta d\varphi^2) \ . 
\end{eqnarray}
We use $\epsilon$ as the control parameter in the linear expansion. 
The complex quantity $\omega=\omega_R+i\omega_I$   
corresponds to the sought-after eigenvalue of the respective quasinormal mode.
Its real part is the oscillation frequency; its imaginary part is the damping rate.

Inserting the Ansatz (\ref{stab3})-(\ref{metric_l0_pert})
into the general set of field equations 
(\ref{Eins})-(\ref{Klein}) 
and utilizing the set of equations for the background functions (\ref{stateq}) 
leads to the Master equation for the spherical perturbations.
This is a single Schr\"odinger-like ODE
for the function $Z=r\phi_1$:
\begin{eqnarray}
\frac{d^2Z}{d(R^*)^2} &=& (U_0(r) - \omega^2)Z \ , 
\end{eqnarray}
which involves the tortoise coordinate $R^*$, where
\begin{eqnarray}
\partial_r R^* &=& \frac{1}{\sqrt{g(1-2m/r)}} \ ,
\end{eqnarray}
and the potential $U_0$, reading
\begin{eqnarray}
U_0(r)&=& 
\frac{r-2m}{2r^5 f(\phi_0) e^{2\delta}}
\left[  \left(Q^2-r^2 f(\phi_0) \right)  \left(r\partial_r\phi_0\right)^2
-4\frac{\dot f(\phi_0)}{f(\phi_0)} Q^2 r \partial_r \phi_0 
\right. \nonumber \\
&+& \left. 
2Q^2 \left(2\left(\frac{\dot f(\phi_0)}{f(\phi_0)}\right)^2-1  \right)
+4rm f(\phi_0) \right] \ .
\end{eqnarray}

To determine the quasinormal modes, we need to impose
an adequate set of boundary conditions.
As the perturbation propages toward infinity, 
we have to impose an outgoing wave behavior,
i.e., 
\begin{eqnarray}
Z =A_{\phi}^{+}e^{i\omega R^*} \left(1+\frac{iM}{2\omega}\frac{1}{r^2} + O(r^{-3})\right) \ 
\label{zout} \, ,
\end{eqnarray}
for $r\to \infty$.
As the perturbation propagates toward the horizon,
we have to impose an ingoing wave behavior,
i.e., 
\begin{eqnarray}
Z &=&A_{\phi}^{-}e^{-i\omega R^*} 
 \\  &&
\left( 1 + 2r_H f(\phi_H)
\frac{ \left(i \left(\frac{\dot f(\phi_H)}{f(\phi_H)}\right)^2
                +e^{\delta_H}\omega r_H  \right) Q^2  
                -f(\phi_H) e^{\delta_H}\omega r_H^3}
    {\left( f(\phi_H)r_H^2-Q^2 \right) 
    \left( -iQ^2 + f(\phi_H) r_H^2
     (i+2r_He^{\delta_H}\omega) \right)} (r-r_H) 
+ O((r-r_H)^2) \right) \, , \nonumber
\label{zin}
\end{eqnarray}
for $r \to r_H$.
In these expansions, the quantities $A_\phi^{\pm}$ denote arbitrary amplitudes
for the perturbation, while all other terms are 
fixed by the background solution.

\subsection{Axial perturbations}

When considering axial perturbations, the scalar field does not get affected
due to symmetry. Here, only the electromagnetic field and the metric enter.
An appropriate Ansatz for the electromagnetic field involves
the perturbation function $W_2$:
\begin{eqnarray}
\label{em_axial_pert}
A= a_0(r) dt -\epsilon W_2(r)e^{-i\omega t} \frac{\partial_{\varphi}Y_{lm}(\theta,\varphi)}{\sin{\theta}}d\theta
+ \epsilon W_2(r)e^{-i\omega t} \sin{\theta}{\partial_{\theta}Y_{lm}(\theta,\varphi)} d\varphi \ , 
\end{eqnarray}
where $Y_{lm}$ are the standard spherical harmonics. 
The corresponding Ansatz for the metric
\begin{eqnarray}
\label{metric_axial_pert}
ds^2&=&-g(r) dt^2 
+ \frac{1}{1-2m(r)/r}dr^2 +r^2(d\theta^2+\sin^2 \theta d\varphi^2)
\nonumber \\
 & & + 2 \epsilon h_0(r)e^{-i\omega t}\frac{\partial_{\varphi}Y_{lm}(\theta,\varphi)}{\sin{\theta}} dtd\theta +
2 \epsilon h_0(r)e^{-i\omega t}\sin{\theta}\partial_{\theta}Y_{lm}(\theta,\varphi) dtd\varphi
\nonumber \\
 & & + 2 \epsilon h_1(r)e^{-i\omega t}\frac{\partial_{\varphi}Y_{lm}(\theta,\varphi)}{\sin{\theta}} drd\theta 
+
2 \epsilon h_1(r)e^{-i\omega t}\sin{\theta}\partial_{\theta}Y_{lm}(\theta,\varphi) drd\varphi \, ,
\end{eqnarray}
introduces the perturbation functions $h_0$ and $h_1$.

Inserting this Ansatz into the field equations 
(\ref{Eins})-(\ref{Klein})
leads to a set of coupled differential equations,
which is shown in the Appendix.
It involves first order equations for 
the metric perturbation functions $h_0$ and $h_1$,
and a second order equation for 
the electromagnetic perturbation function $W_2$.
This system can be put into the form
\begin{eqnarray}
\partial_r \Psi_{A} = M_{A} \Psi_{A} \ ,
\end{eqnarray}
where $\Psi_A$ denotes the perturbation functions in the form
\begin{eqnarray}
\Psi_A = \left[ \begin{array}{c}
h_0
\\ 
h_1
\\ 
W_2
\\ 
\partial_r W_2
\end{array} 
\right] \ ,
\end{eqnarray}
and $M_{A}$ denotes a $4\times 4$ matrix
which contains the background functions,
the angular number $l$, 
and the eigenvalue $\omega$ of the quasinormal mode.
To determine the axial quasinormal modes,
as for the radial ones, we have to impose the proper set
of boundary conditions at infinity (outgoing) and at
the horizon (ingoing).
These can be found in the Appendix.

\subsection{Polar perturbations}

Since the radial perturbations are a set of polar perturbations ($l=0$),
it is clear that the general polar perturbations involve again all the fields.
For the scalar field, we introduce the perturbation function $\phi_1$
and the Ansatz
\begin{eqnarray}
\label{s_polar}
\phi&=&\phi_0(r) + \epsilon e^{-i \omega t} \phi_1(r) Y_{lm}(\theta,\varphi) \ .
\end{eqnarray} 
For the electromagnetic field, we employ the perturbation functions $a_1$, $W_1$ and $V_1$
and the Ansatz
\begin{eqnarray}
\label{A_polar}
A&=& ( a_0(r)+\epsilon e^{-i \omega t} a_{1}(r) Y_{lm}(\theta,\varphi)) dt 
+ \epsilon W_1(r) e^{-i \omega t} Y_{lm}(\theta,\varphi) dr
 \nonumber \\
& &+ \epsilon V_1(r) e^{-i \omega t} \partial_{\theta}Y_{lm}(\theta,\varphi) d\theta + \epsilon V_1(r) e^{-i \omega t} \partial_{\varphi}Y_{lm}(\theta,\varphi) d\varphi \ .
\end{eqnarray} 
Finally, for the metric, we introduce the perturbation functions $N$, $H_1$, $L$ and $T$,
and employ the Ansatz 
\begin{eqnarray}
\label{metric_polar_pert}
ds^2&=&-g(r) (1+\epsilon e^{-i \omega t} N(r) Y_{lm}(\theta,\varphi)) dt^2 
- 2\epsilon e^{-i\omega t} H_1(r) Y_{lm}(\theta,\varphi) dtdr
+ \frac{1-\epsilon e^{-i \omega t} L(r)Y_{lm}(\theta,\varphi)}{1-2m(r)/r}dr^2
\nonumber \\
& &+(r^2-2\epsilon e^{-i \omega t} T(r)Y_{lm}(\theta,\varphi))(d\theta^2+\sin^2 \theta d\varphi^2) \ .
\end{eqnarray}

Inserting this Ansatz into the field equations 
(\ref{Eins})-(\ref{Klein}) 
then again results in a set of coupled differential equations,
which are shown in the Appendix.
In fact, this system of equations can be simplified
when one introduces the new functions $F_0$, $F_1$ and $F_2$,
that are defined in terms of the perturbation functions $W_1$, $V_1$ and $a_1$
(as discussed in the Appendix).
The resulting Master equations
can again be put in vectorial form,
\begin{eqnarray}
\partial_r \Psi_P = M_P \Psi_P \ .
\label{polar_master_Eq}
\end{eqnarray} 
The vector now contains 6 components:
\begin{eqnarray}
\Psi_P = \left[ \begin{array}{c}
H_1
\\ 
T
\\ 
F_0
\\ 
F_1
\\
\phi_1
\\
\partial_r \phi_1
\end{array} 
\right] \ .
\end{eqnarray}
The $6 \times 6$  matrix $M_P$
depends again on the background functions,
the angular number $l$ and the eigenvalue $\omega$ of the quasinormal mode.

We remark that
for a RN black hole, the equations for the scalar perturbations are decoupled.
However, the equations for the electromagnetic perturbations couple with the equations
for the metric perturbations.
When the charged black hole also carries scalar hair,
all equations are coupled to each other.

Again, we must impose the proper boundary conditions
at infinity (outgoing) and at the horizon (ingoing)
to determine the quasinormal modes.
As in the axial case, these are shown in the Appendix.

\section{Quasinormal mode spectrum}

In the following, we briefly discuss the method used to extract the quasinormal mode
spectrum of the EMs black holes and recall the nomenclature for the modes.
Then, we present our numerical results for the $l=0$, $l=1$ and $l=2$ cases
and discuss isospectrality breaking.

\subsection{Numerics and nomenclature}

We calculate the quasinormal modes of the black holes on all three branches,
bald, cold and hot,
making sure that
the respective spectra match,
when the background solutions get close.
Since the RN quasinormal modes are well-known
(see e.g., \cite{Onozawa:1995vu,Andersson:1996xw,Kokkotas:1988fm,Leaver:1990zz}), 
this provides an independent check of the numerics used, as well as a first guess for the spectrum of the cold branch, which is typically close to the RN branch for large values of the electric charge.
All calculations are performed for a coupling constant $\alpha=200$.

In a first step, we obtain the background solutions with high precision,
solving numerically the set of  ODEs (\ref{stateq}),
subject to the boundary conditions following from the expansions
at infinity (\ref{inf_static}) and at the horizon (\ref{hor_static}).
For this purpose, we employ the solver COLSYS \cite{Ascher:1979iha},
which uses a collocation method for boundary-value ODEs together with 
a damped  Newton method of quasi-linearization. 
The problem is linearized and solved at each iteration step, 
employing a spline collocation at Gaussian points. 
The solver features an adaptive mesh selection procedure, 
refining the grid until the required accuracy is reached.

Once the background solutions are known, we follow the
procedure that is analogous to the one we have used before
to calculate the quasinormal modes of hairy black holes
\cite{Blazquez-Salcedo:2020rhf,Blazquez-Salcedo:2020caw,Blazquez-Salcedo:2016enn,Blazquez-Salcedo:2017txk,Blazquez-Salcedo:2019nwd}. 
We split the space-time into two regions,
the inner region $r_H+\epsilon_H \le r \le r_J$, 
and the outer region $r_J \le r \le r_\infty$. 
In the inner region, we impose the respective ingoing wave behavior; 
in the outer region, we impose the respective outgoing wave behavior.
Then, we calculate sets of linearly independent solutions numerically
and match them at the common border $r_J$ of the two regions.
The eigenvalue $\omega$ of the quasinormal modes is found
when the functions and their derivatives are continuous at the matching point $r_J$.

We follow a common nomenclature for quasinormal modes, that is used
when scalar and electromagnetic fields are coupled in the background solutions.
Without such a coupling in the background solutions,
one would simply obtain scalar or electromagnetic or gravitational modes
by solving the respective scalar, electromagnetic or gravitational perturbation equations, 
since the different types of perturbations would not be coupled to each other.
However, when all fields are already present in the background solutions,
the different types of perturbations couple.
By taking the respective charges, $Q_s$ and $Q$, to zero, the perturbations
decouple again.
Therefore, we employ a nomenclature that reveals this decoupling limit.
\begin{itemize}
\item[i.]
Modes that are connected with purely scalar perturbations 
are called scalar-led modes. 
Typically they have dominant amplitude $A_\phi^{\pm}$.
\item[ii.]
Modes that are connected with purely electromagnetic perturbations
are called EM-led modes. 
Typically they have dominant amplitude $A_F^{\pm}$.
\item[iii.]
Modes that are connected with purely gravitational perturbations 
are called grav-led modes. 
Typically they have dominant amplitude $A_g^{\pm}$.
\end{itemize}
Scalar-led modes exist for $l \ge 0$, EM-led modes for $l \ge 1$
and grav-led for $l \ge 2$.
In the following, we present our results for the quasinormal modes 
for the cases $l=0$, $1$ and $2$, successively.

\subsection{Spectrum of $l=0$ quasinormal modes}

\begin{figure}[t]
	\centering
	\includegraphics[width=0.35\linewidth,angle=-90]{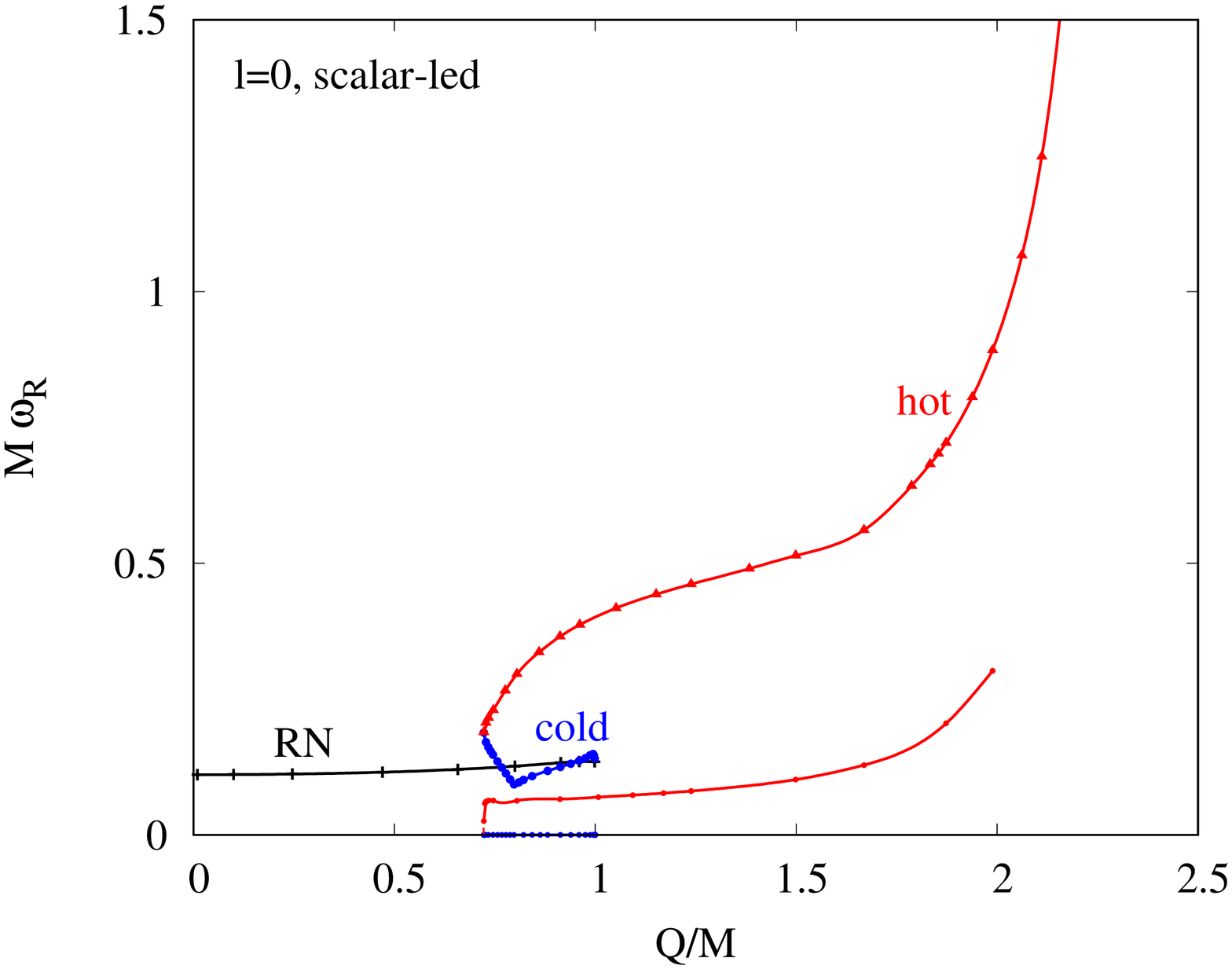}
      \includegraphics[width=0.35\linewidth,angle=-90]{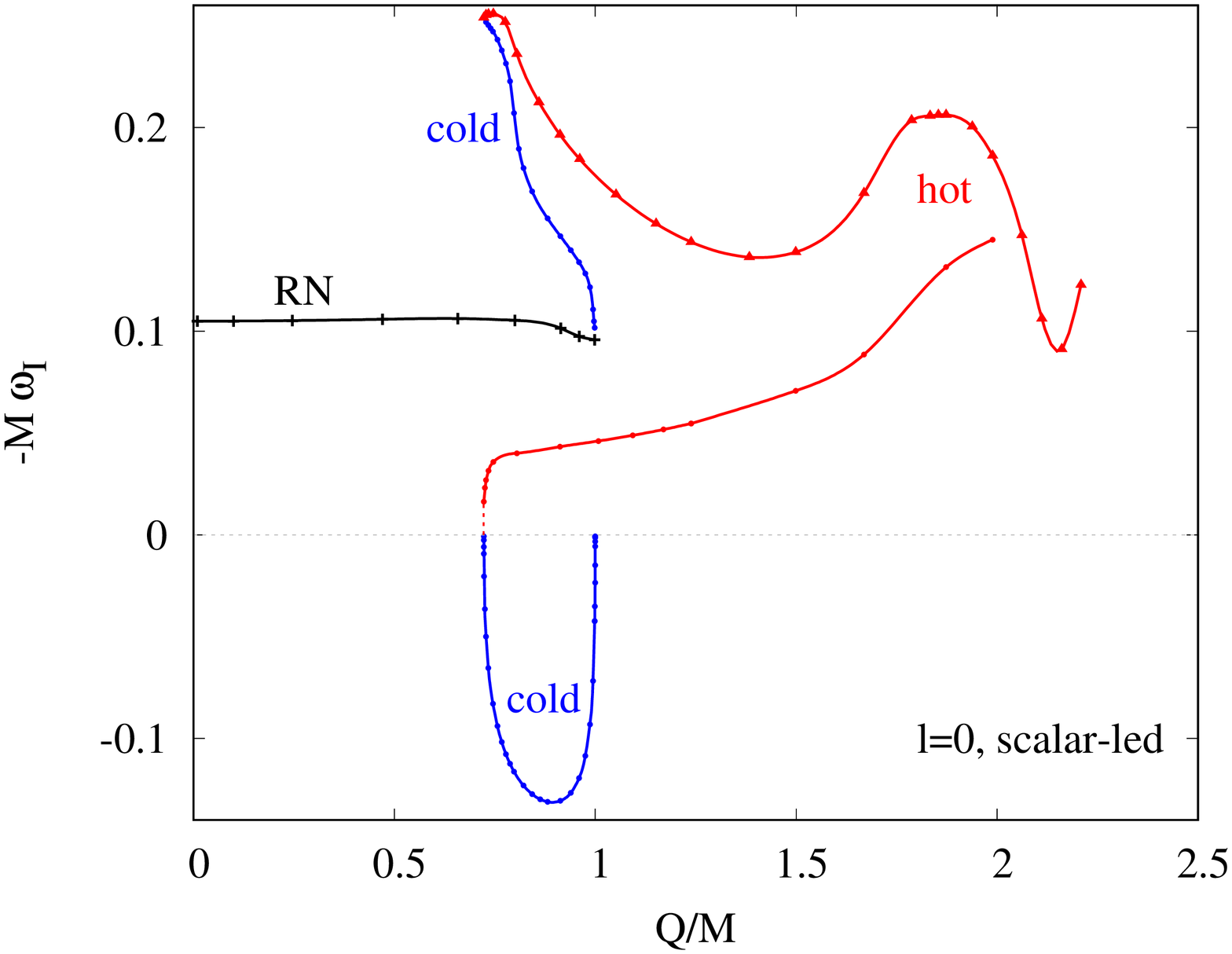}
	\caption{Scalar-led modes for $l=0$ perturbations: 
	scaled real part $\omega_R/M$ (a) and scaled imaginary part $-\omega_I/M$ (b) 
       vs. reduced charge $q$
       for RN (black), and cold (blue) and hot (red) scalarized black holes ($\alpha=200$).}
	\label{fig:l0_scalar_R}
\end{figure}

The $l=0$ perturbations are most interesting, since they 
are the ones that discriminate between stable and unstable EMs black hole solutions.
In particular, they feature an unstable mode on the cold black hole branch,
whereas the bald and the hot black hole branches possess only stable modes,
as our study shows.

We exhibit the lowest scalar-led $l=0$ modes vs. the charge to mass ratio $q$
in Fig.~\ref{fig:l0_scalar_R},
showing the scaled real part of the frequency, $\omega_R/M$, in Fig.~\ref{fig:l0_scalar_R}(a)
and the scaled imaginary part $-\omega_I/M$ in Fig.~\ref{fig:l0_scalar_R}(b).
A positive imaginary part signals instability, whereas a negative imaginary part
yields the damping time of the mode.
Following the color coding of Fig.~\ref{fig:bh},
the bald RN branch is shown in black, the cold EMs branch in blue, and the hot EMs branch in red. 

As seen in Fig.~\ref{fig:l0_scalar_R},
the fundamental RN branch has only little dependence on $q$,
deviating only slightly from the Schwarzschild mode all the way up to the 
extremal black hole. The full RN branch is also free of unstable modes.

The cold EMs branch features an unstable scalar-led mode
throughout its domain of existence.
The mode tend to vanish for the largest charge to mass ratio of the cold branch, $q=1$.
The purely imaginary frequency grows as $q$ decreases. At a certain value of $q$ it reaches a maximum, from where it decreases again to the end point of the branch,
where it reaches a zero mode.
Here the bifurcation with the hot branch is encountered, and thus the minimal
charge to mass ratio $q_{\text{min}}$ of both EMs branches is reached.
Continuity then requires that at $q_{\text{min}}$, also the hot EMs branch
has a zero mode.

Besides the unstable mode, the cold branch features a stable
scalar-led mode, which is close to the scalar-led mode of the RN black hole.
Along this branch, from $q=1$ to $q_{\text{min}}$ the frequency $\omega_R/M$
first decreases slowly and then increases, becoming larger than the frequency
of the RN branch, while the damping rate $|\omega_I/M|$ increases monotonically.

The corresponding stable scalar-led modes of the hot EMs branch extend from $q_{\text{min}}$
to $q_{\text{max}}$, where an extremal singular EMs solution is reached.
The fundamental branch starts at the zero mode at $q_{\text{min}}$,
from where both the frequency $\omega_R/M$ and the damping rate $|\omega_I/M|$
rise at first almost vertically. They continue to rise monotonically,
reaching final values above those of the extremal RN branch.
The first overtone branch starts at the stable scalar-led mode at the bifurcation point.
Its frequency $\omega_R/M$ rises also monotonically, but its damping rate $|\omega_I/M|$
exhibits an overall but not monotonic decrease.

The reason, we exhibit not only the fundamental stable mode for the hot EMs branch 
but also the first overtone is to 
demonstrate continuity of the modes at the bifurcation with the cold EMs branch.
The zero mode of the cold EMs branch turns into the fundamental mode of the
hot EMs branch, whereas the fundamental mode of the RN branch can be followed
via the first stable mode of the cold EMs branch to the first stable overtone
of the hot EMs branch.

Of course, all three classical branches feature sequences of (further) overtones,
not studied here in detail. In the RN case, they are well known. 
There, the rapidly damped modes possess several peculiar features. For instance,
the higher modes of the non-extremal black holes have been observed to spiral towards the modes
of the extremal black hole with increasing $q$
\cite{Onozawa:1995vu,Andersson:1996xw}.

\subsection{Spectrum of $l=1$ quasinormal modes}

\begin{figure}[t]
	\centering
	\includegraphics[width=0.35\linewidth,angle=-90]{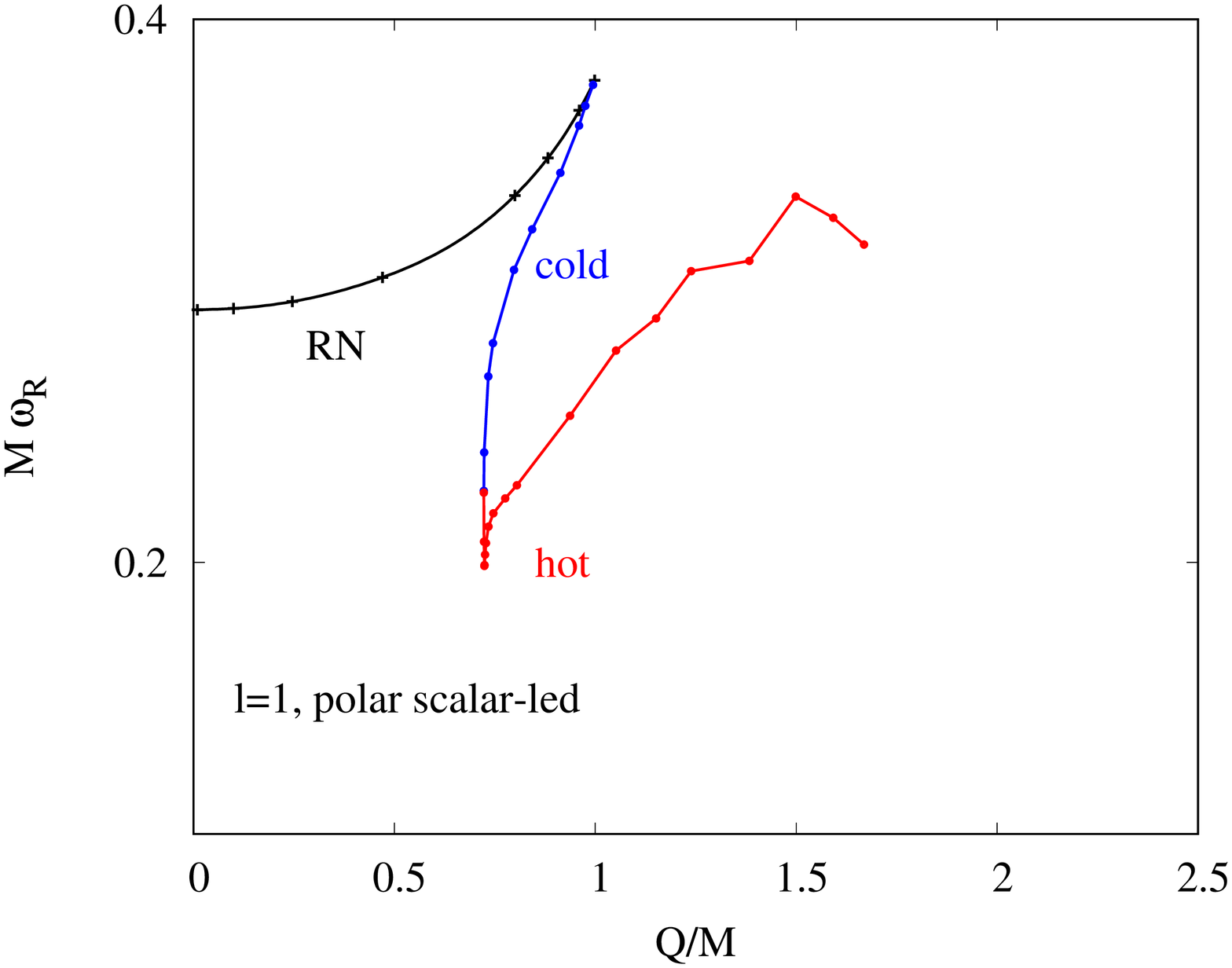}
      \includegraphics[width=0.35\linewidth,angle=-90]{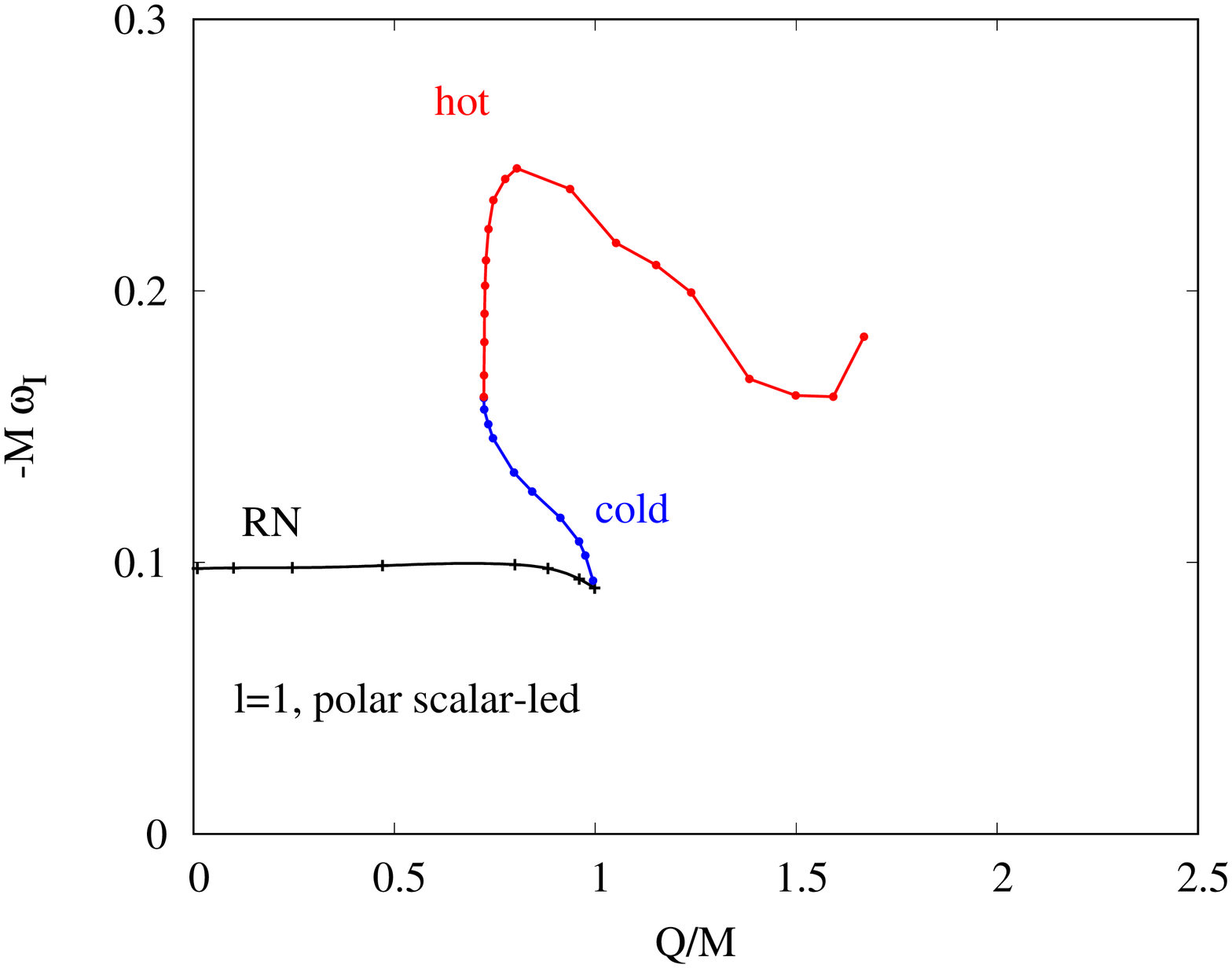}
	\caption{Scalar-led modes for polar $l=1$ perturbations: 
	scaled frequency $\omega_R/M$ (a) and scaled damping rate $-\omega_I/M$ (b) 
       vs. reduced charge $q$
       for RN (black), and cold (blue) and hot (red) scalarized black holes ($\alpha=200$).}
	\label{fig:l1_scalar_R}
\end{figure}
\begin{figure}[h!]
	\centering
	\centering
	\includegraphics[width=0.35\linewidth,angle=-90]{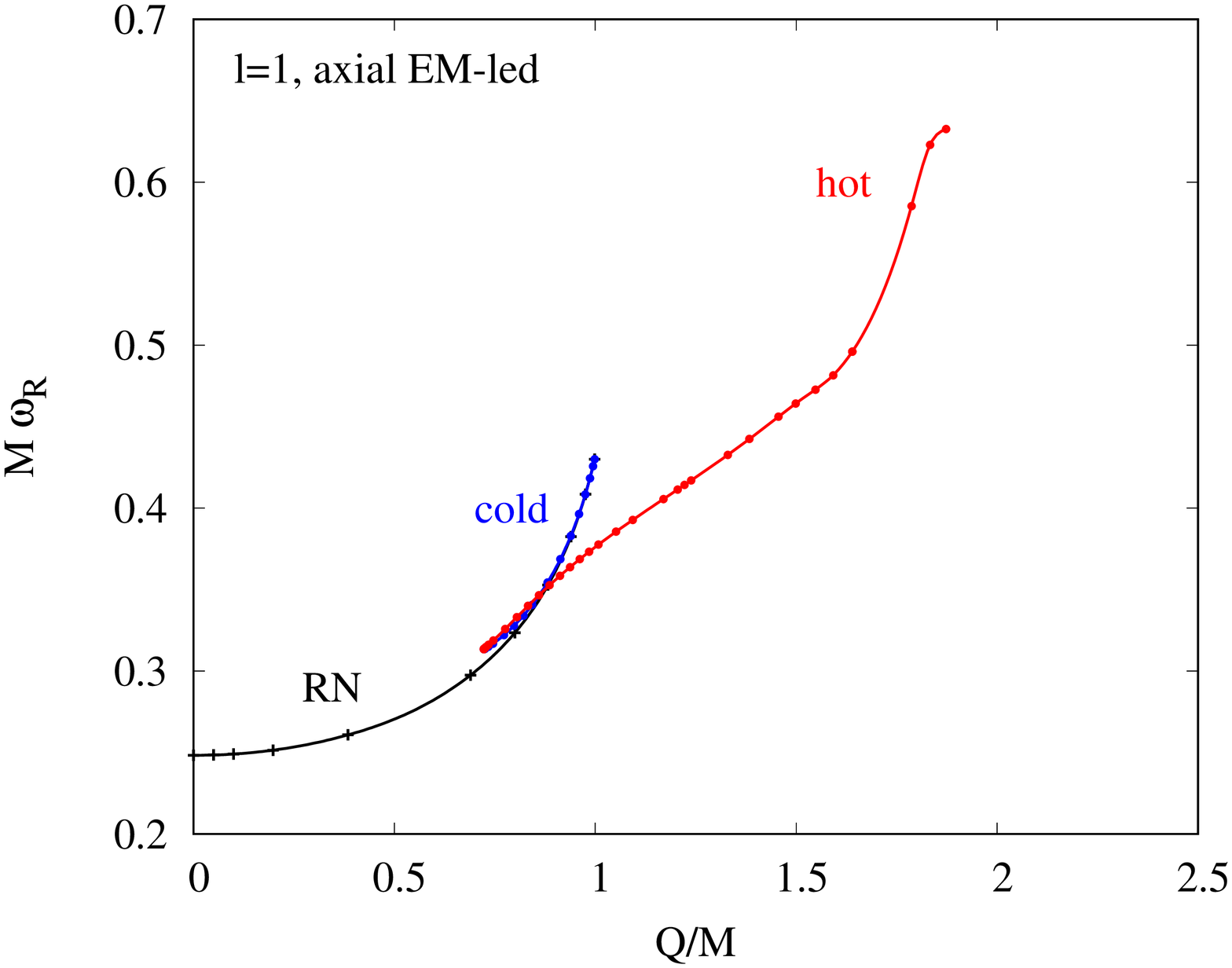}
      \includegraphics[width=0.35\linewidth,angle=-90]{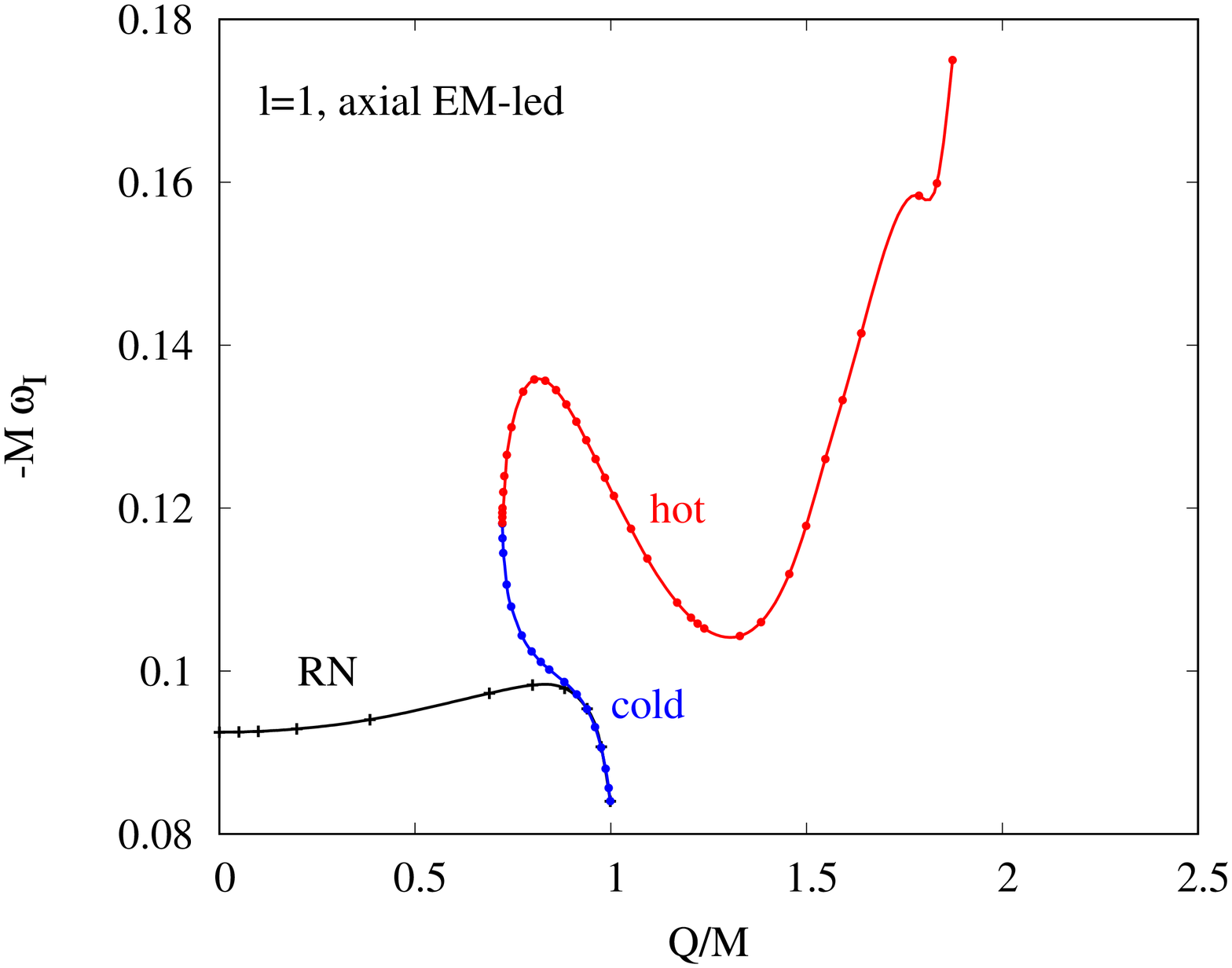}
      \includegraphics[width=0.35\linewidth,angle=-90]{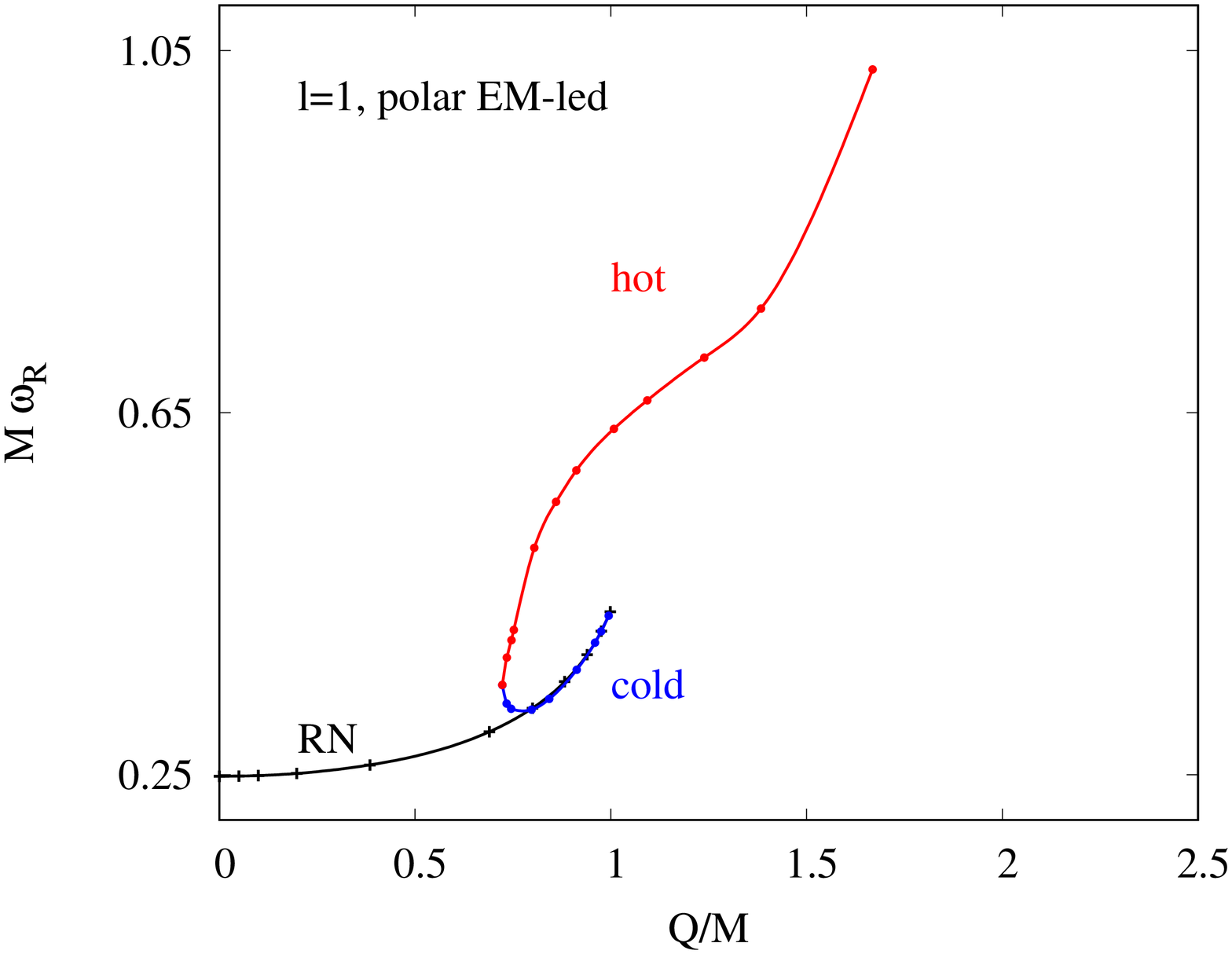}
      \includegraphics[width=0.35\linewidth,angle=-90]{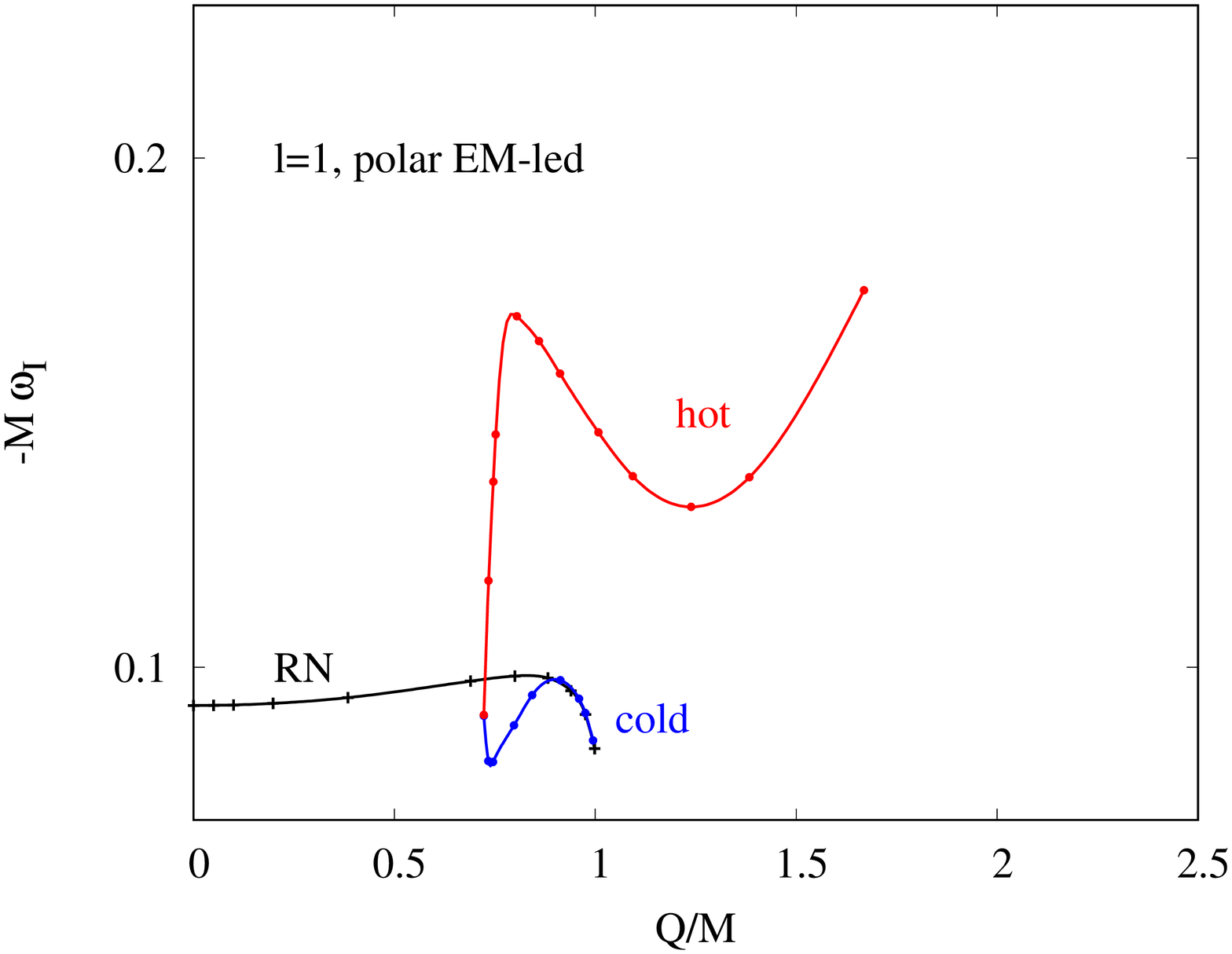}
	\caption{EM-led modes for axial (a)-(b) and polar (c)-(d) $l=1$ perturbations: 
	scaled frequency $\omega_R/M$ (a), (c) and scaled damping rate $-\omega_I/M$ (b), (d)
       vs. reduced charge $q$
       for RN (black), and cold (blue) and hot (red) scalarized black holes ($\alpha=200$).}
	\label{fig:l1_EMap_R}
\end{figure}

The $l=1$ modes consist of polar scalar-led modes and both axial and polar EM-led modes.
We start the discussion with the scalar-led modes, shown in Fig.~\ref{fig:l1_scalar_R}.
Again, the lowest RN mode changes smoothly with increasing charge to mass ratio $q$.
The frequency $\omega_R/M$ increases monotonically, while the damping rate $|\omega_I/M|$
remains almost constant, showing a slight decrease towards the extremal endpoint $q=1$.

The lowest scalar-led $l=1$ mode of the cold EMs branch changes smoothly 
from the $q=1$ point to the bifurcation point
with the hot EMs branch at $q_{\text{min}}$, exhibiting a monotonic 
decrease of the frequency $\omega_R/M$ and a monotonic increase of 
damping rate $|\omega_I/M|$.
Along the hot EMs branch, the change of the eigenvalue with increasing $q$
is no longer monotonic. As compared to the mode of the extremal RN solution,
the frequency $\omega_R/M$ of the extremal EMs solution is lower
and the damping rate $|\omega_I/M|$ is higher.

In Fig.~\ref{fig:l1_EMap_R} we exhibit the axial and polar EM-led $l=1$ modes.
The RN black holes are known to exhibit isospectrality, i.e., the axial and polar 
EM-led modes are degenerate. The EM-led RN modes have an analogous
$q$-dependence to the scalar-led RN modes, but their absolute values differ
somewhat. 

Starting from the $q=1$ bifurcation point, the axial and polar modes
of the cold EMs branch follow closely the RN modes at first.
The frequencies $\omega_R/M$ of both axial and polar modes are slightly higher
than the RN frequencies, but do not deviate strongly even at 
the bifurcation point $q_{\text{min}}$ with the hot EMs branch.
The damping rates $|\omega_I/M|$ start to deviate from the RN
damping rate earlier, showing an opposite behavior for the axial and polar modes.
For the axial modes, the damping rates increase monotonically,
whereas for the polar modes, a more sinusoidal pattern is seen.

Along the hot EMs branch from $q_{\text{min}} \le q \le q_{\text{max}}$,
the frequencies  $\omega_R/M$ of both axial and polar EM-led modes
rise monotonically, with the polar frequencies rising almost twice
as much compared to the axial ones.
The damping rates $|\omega_I/M|$ change in  a non-monotonic manner
again, first rising from the bifurcation point, and then exhibiting some
oscillation.
At the extremal EMs solution, the damping rates  reach similar values,
corresponding to about twice the extremal RN value.

\subsection{Spectrum of $l=2$ quasinormal modes}

\begin{figure}[t]
	\centering
	\includegraphics[width=0.35\linewidth,angle=-90]{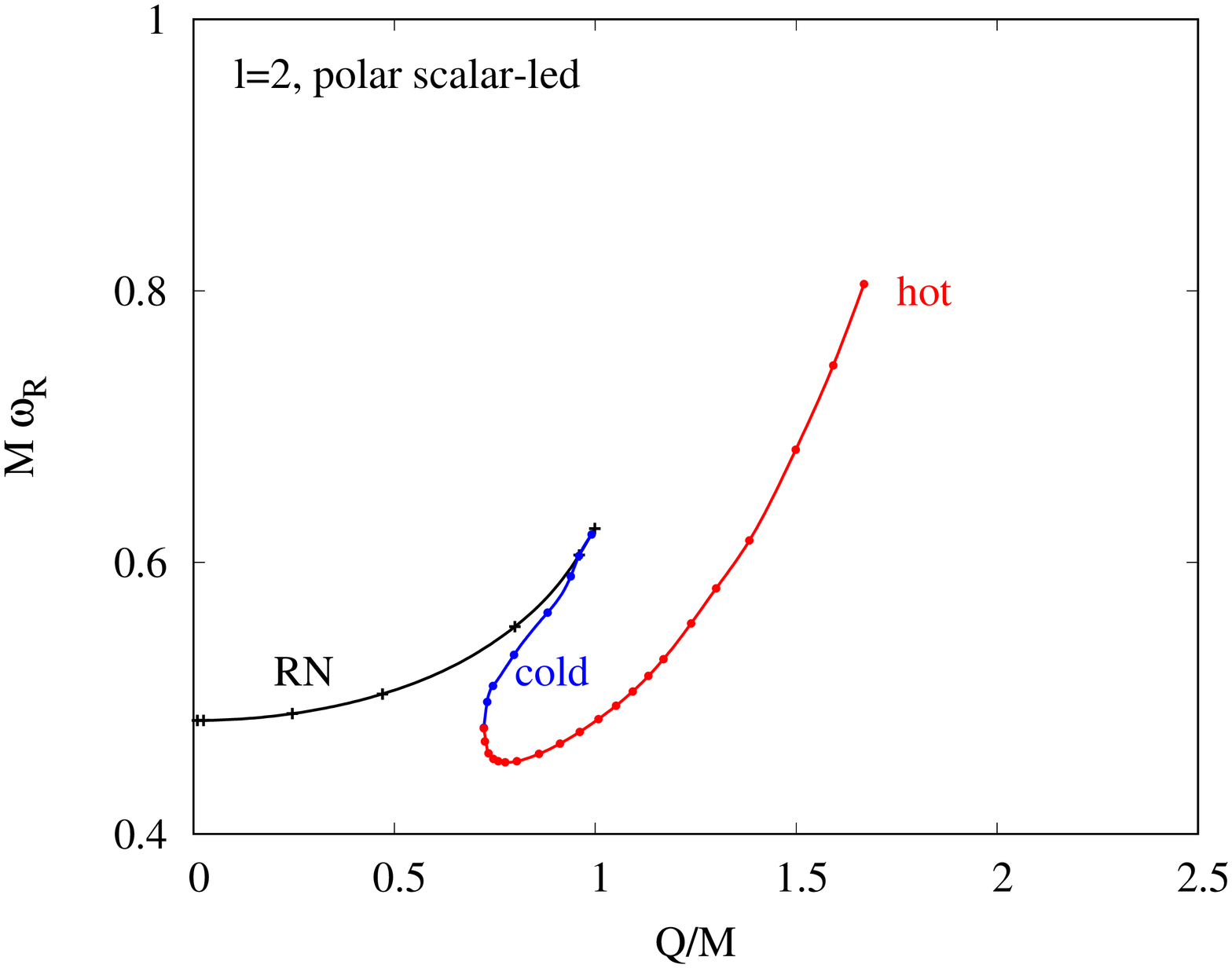}
      \includegraphics[width=0.35\linewidth,angle=-90]{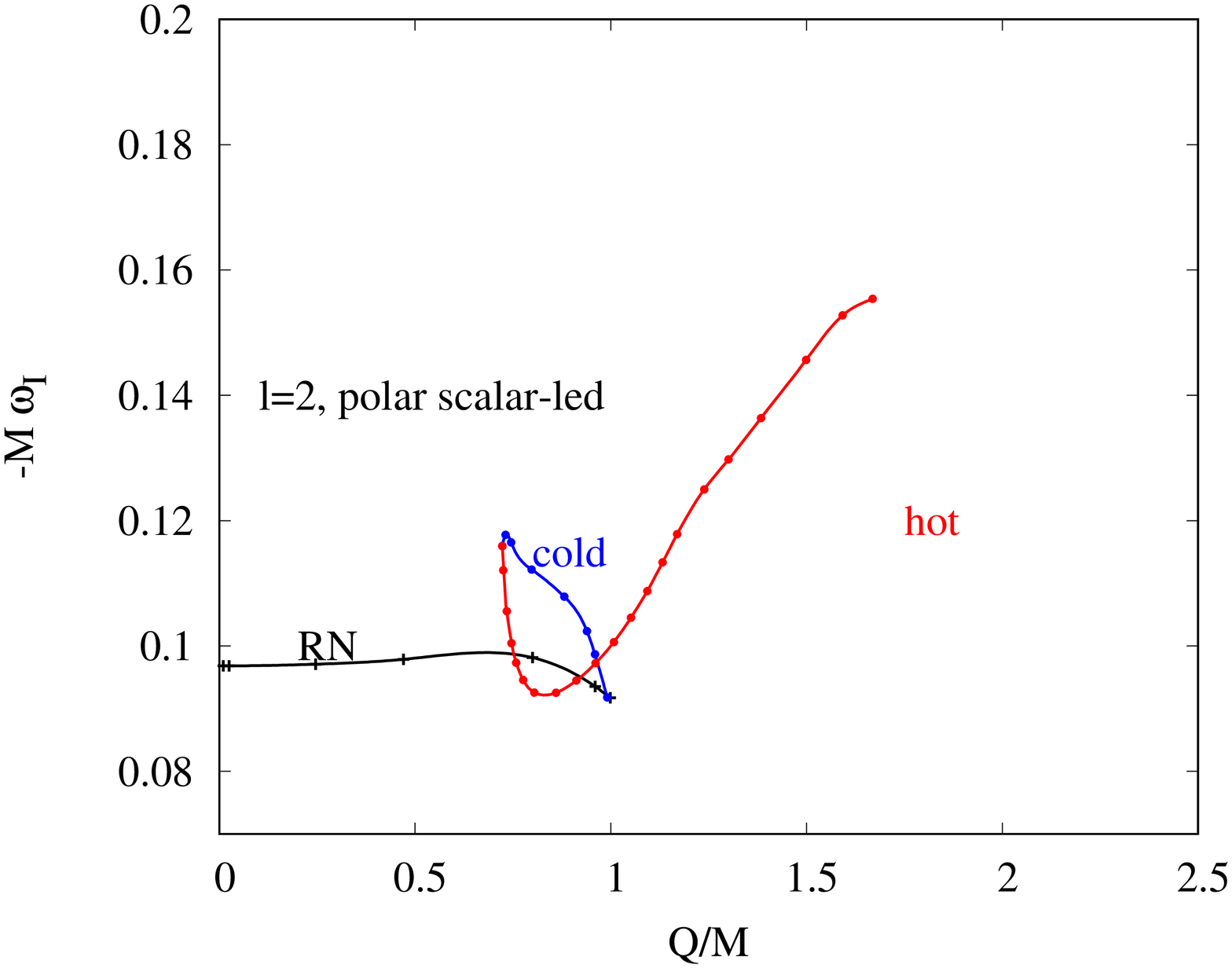}
	\caption{Scalar-led modes for polar $l=2$ perturbations: 
	scaled frequency $\omega_R/M$ (a) and scaled damping rate $-\omega_I/M$ (b) 
       vs. reduced charge $q$
       for RN (black), and cold (blue) and hot (red) scalarized black holes ($\alpha=200$).}
	\label{fig:l2_scalar_R}
\end{figure}
\begin{figure}[h!]
	\centering
	\includegraphics[width=0.35\linewidth,angle=-90]{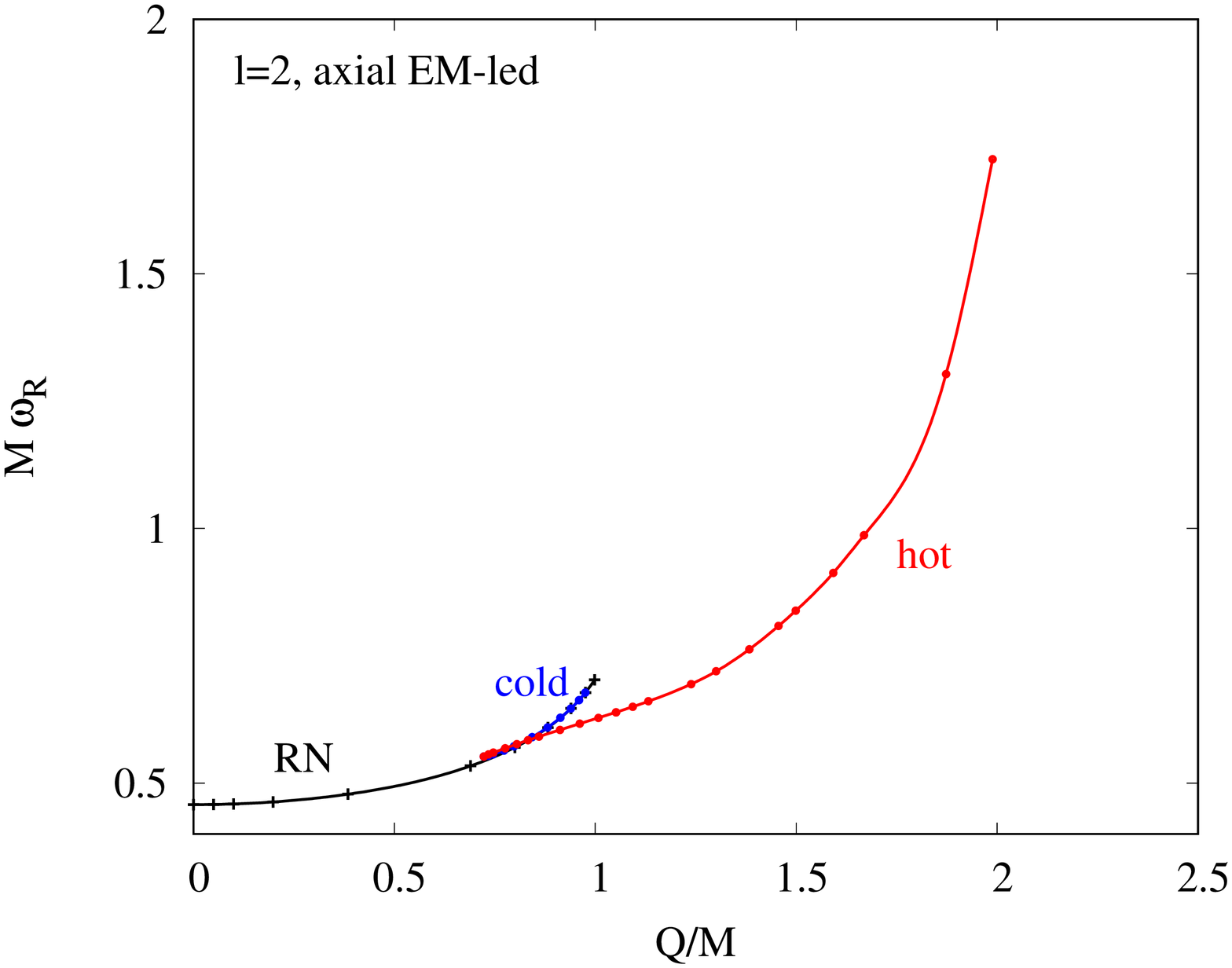}
      \includegraphics[width=0.35\linewidth,angle=-90]{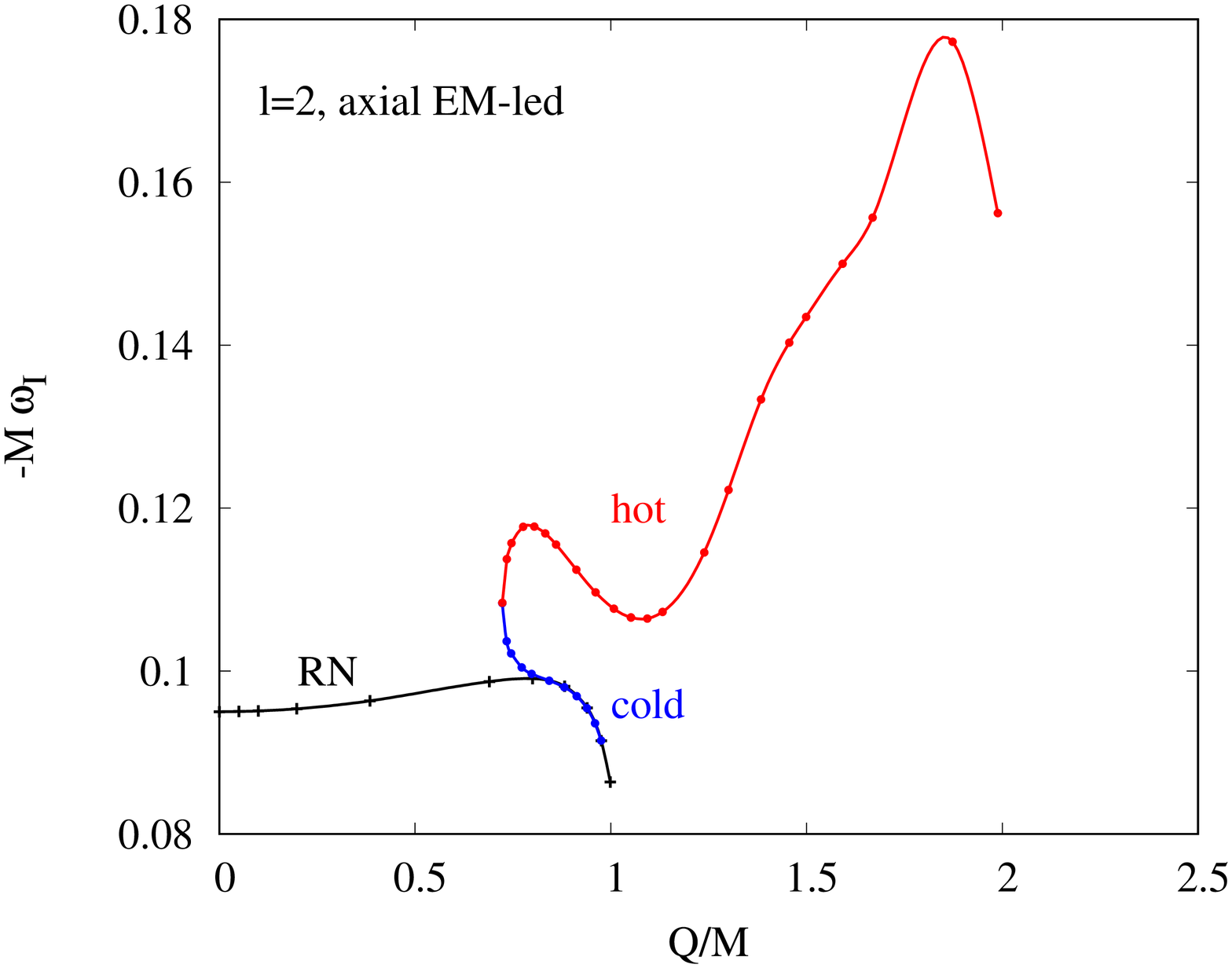}
      \includegraphics[width=0.35\linewidth,angle=-90]{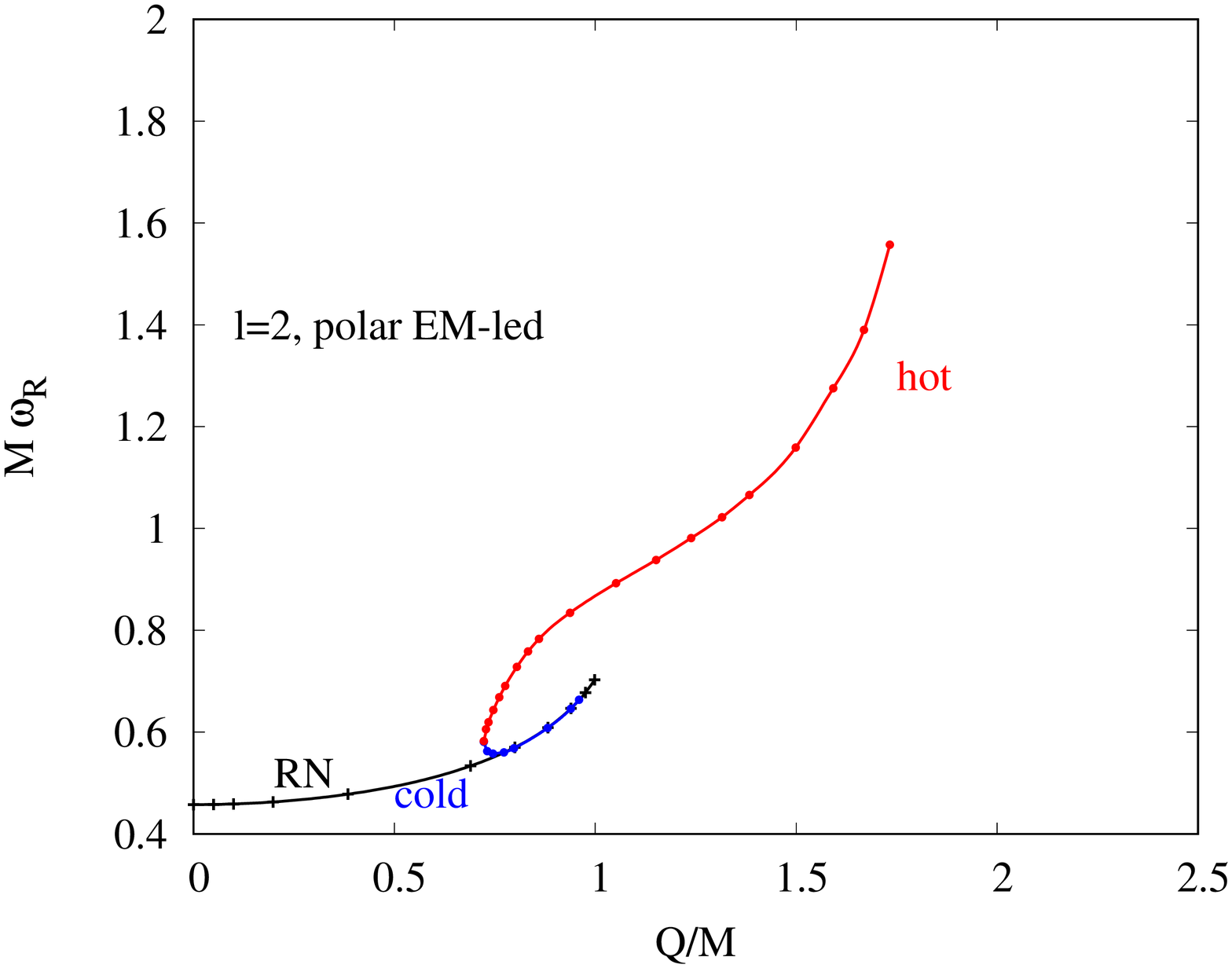}
      \includegraphics[width=0.35\linewidth,angle=-90]{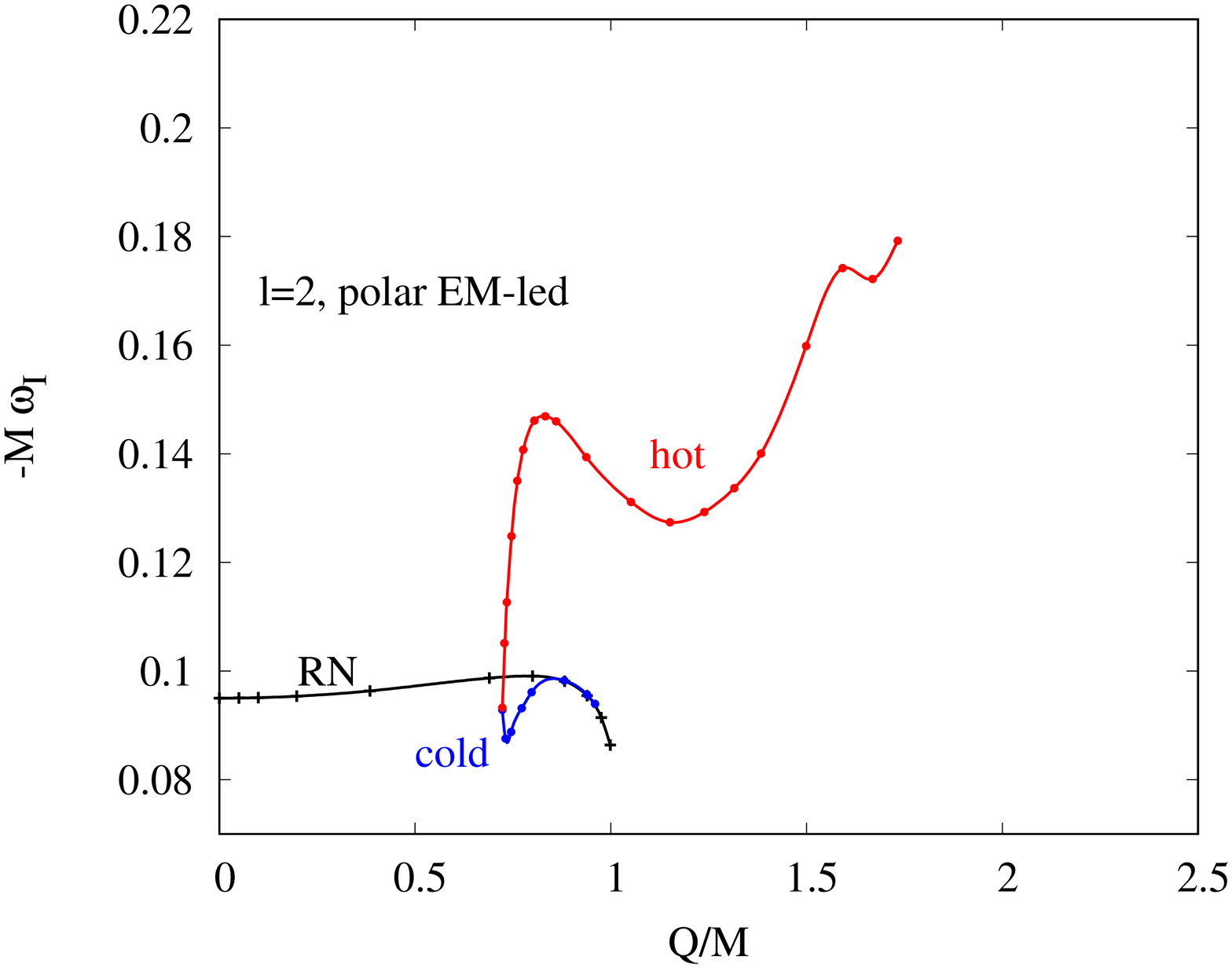}
	\caption{EM-led modes for axial (a)-(b) and polar (c)-(d) $l=2$ perturbations: 
	scaled frequency $\omega_R/M$ (a), (c) and scaled damping rate $-\omega_I/M$ (b), (d)
       vs. reduced charge $q$
       for RN (black), and cold (blue) and hot (red) scalarized black holes ($\alpha=200$).}
	\label{fig:l2_EMap_R}
\end{figure}
\begin{figure}[h!]
	\centering
	\includegraphics[width=0.35\linewidth,angle=-90]{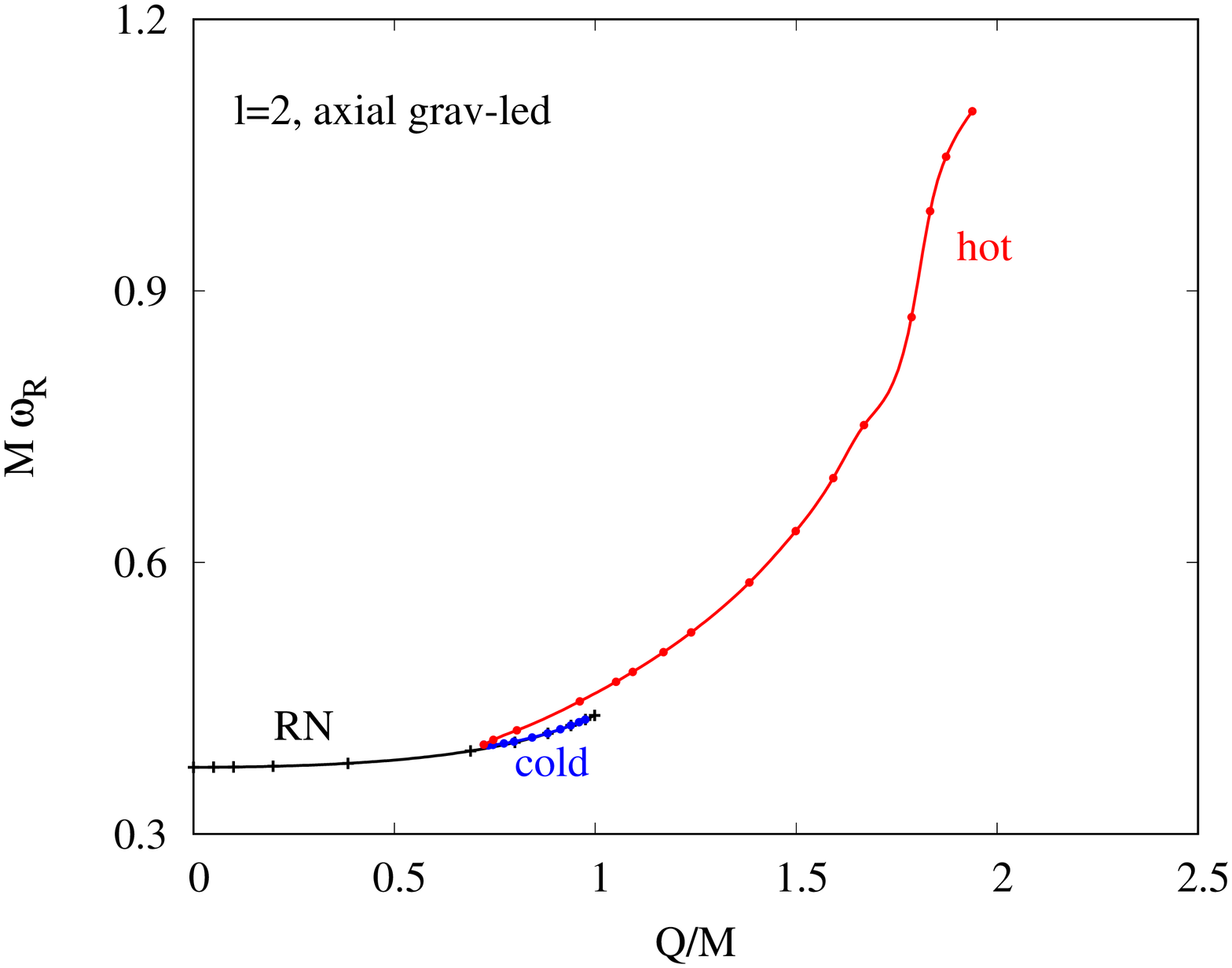}
      \includegraphics[width=0.35\linewidth,angle=-90]{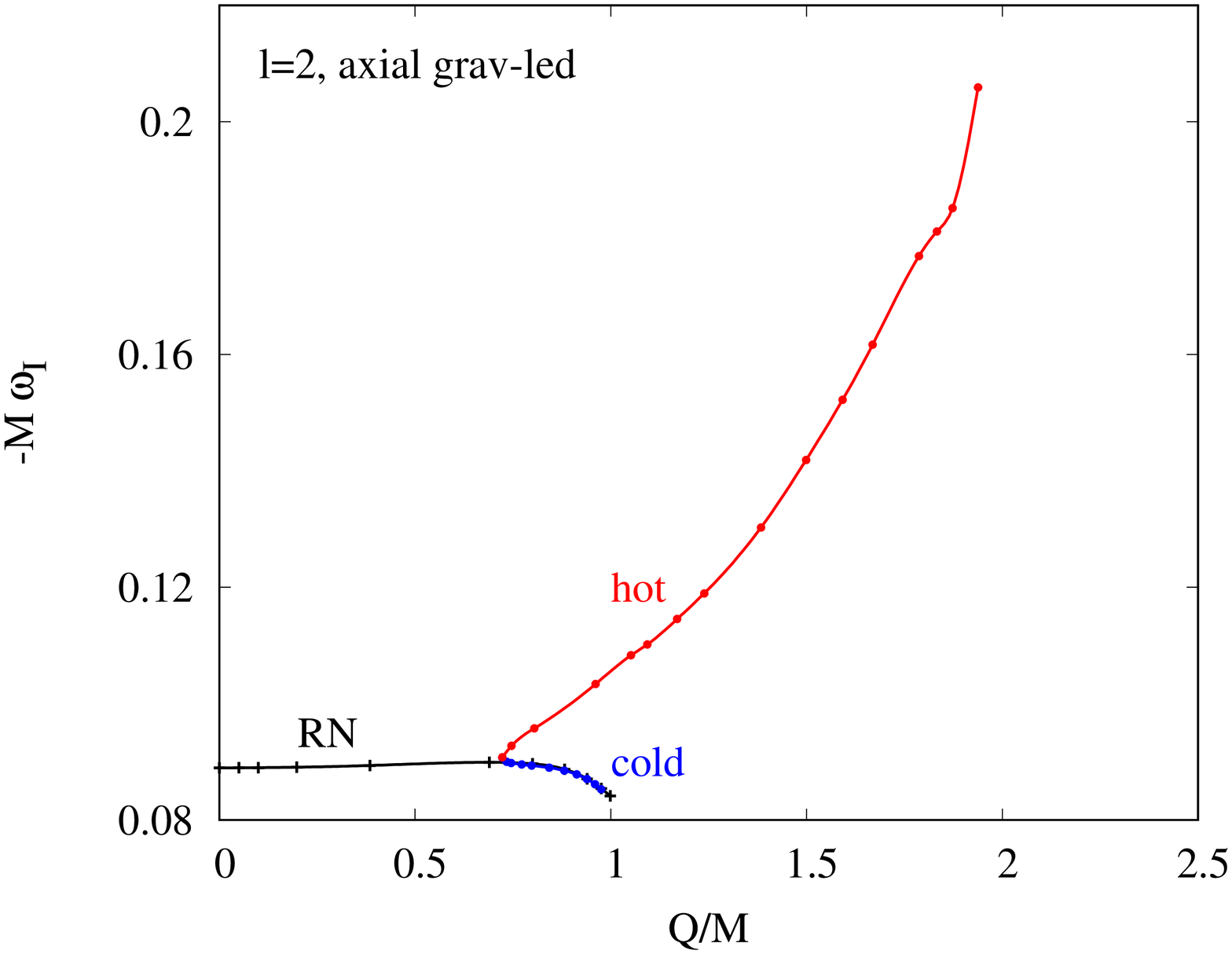}
      \includegraphics[width=0.35\linewidth,angle=-90]{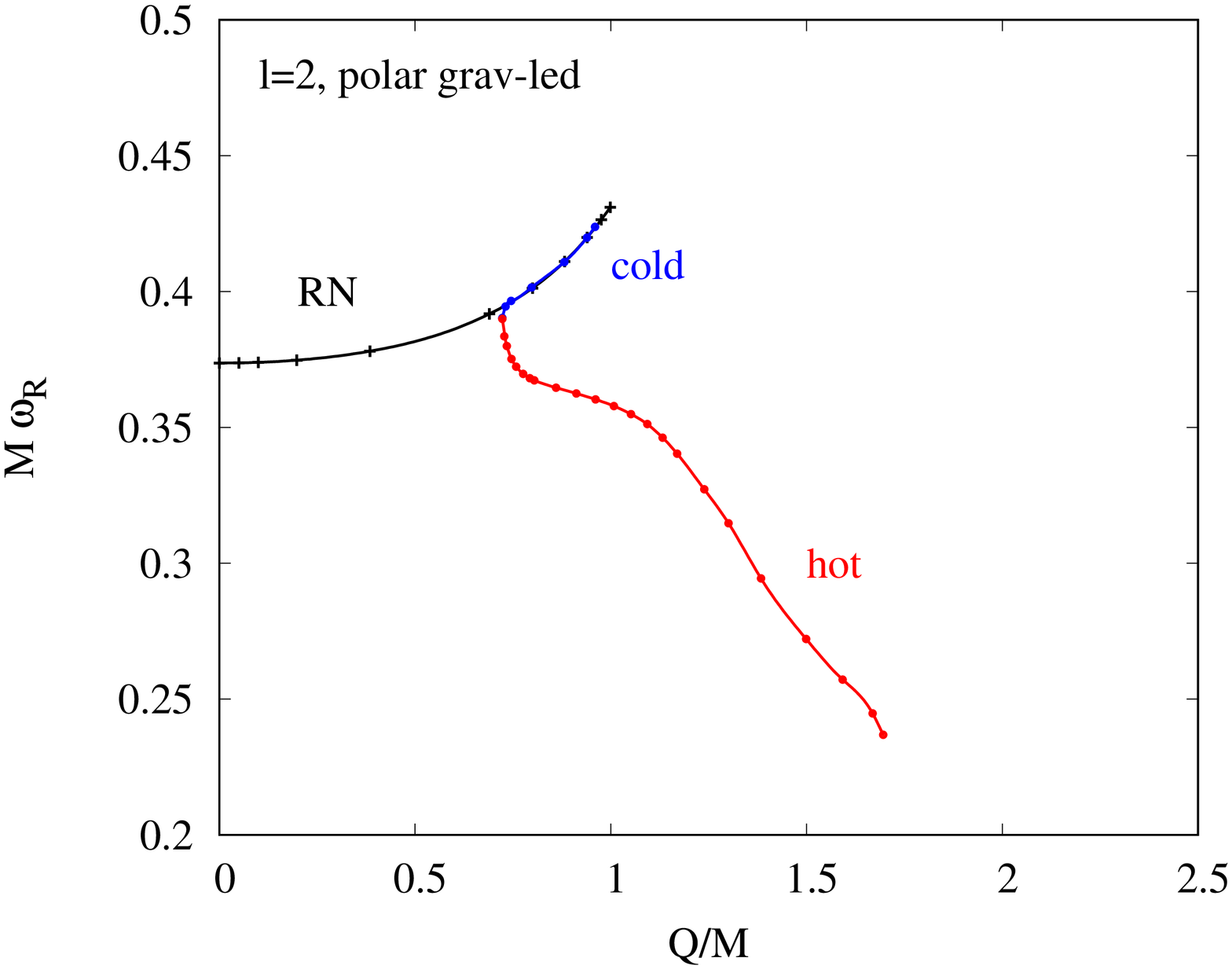}
      \includegraphics[width=0.35\linewidth,angle=-90]{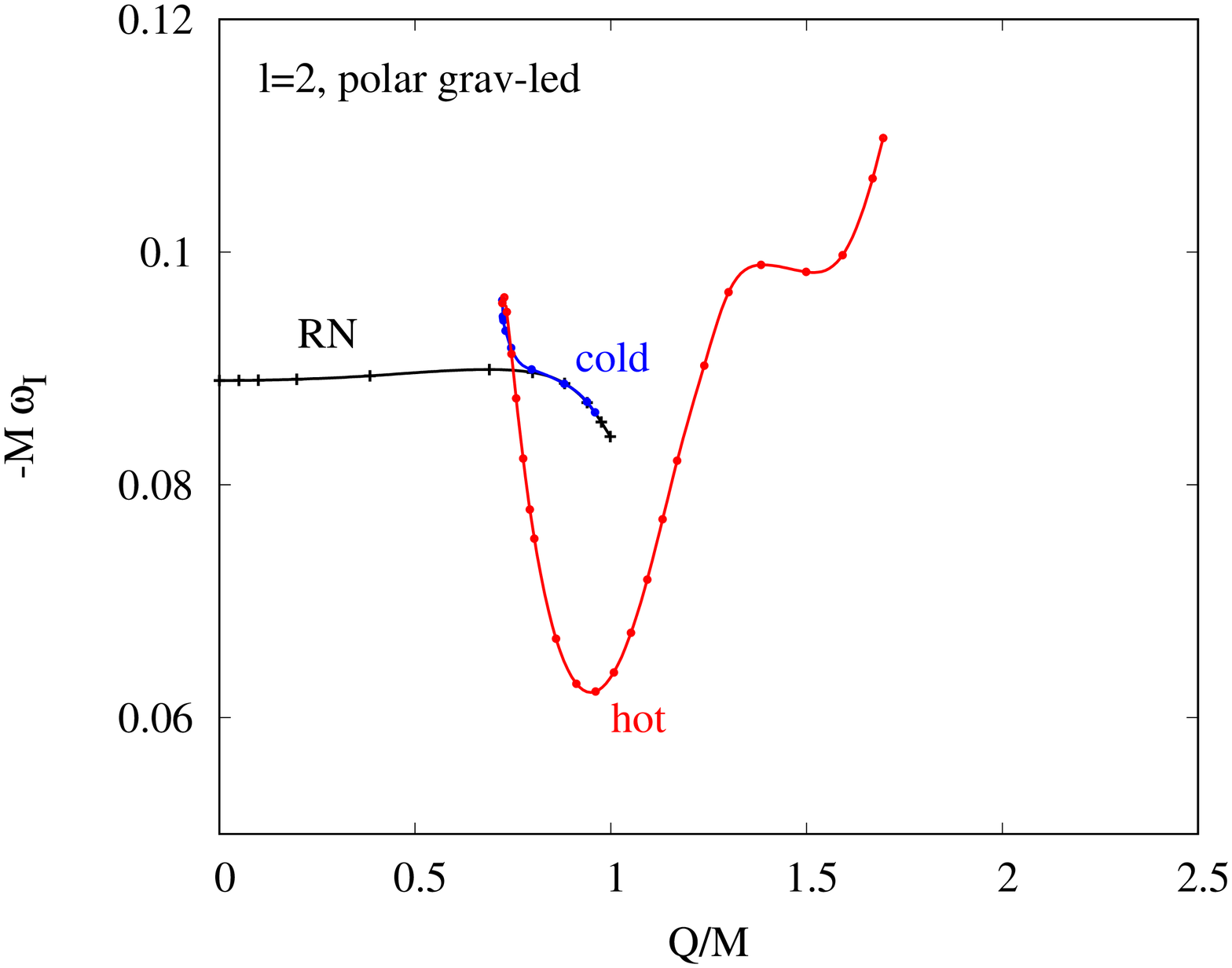}
	\caption{Grav-led modes for axial (a)-(b) and polar (c)-(d) $l=2$ perturbations: 
	scaled frequency $\omega_R/M$ (a), (c) and scaled damping rate $-\omega_I/M$ (b), (d)
       vs. reduced charge $q$
       for RN (black), and cold (blue) and hot (red) scalarized black holes ($\alpha=200$).}
	\label{fig:l2_gravap_R}
\end{figure}

We now turn to the $l=2$ modes, which consist of 
polar scalar-led modes, axial and polar EM-led modes, and additionally
axial and polar grav-led modes. 
Thus, we now have five types of modes for generic EMs black holes.
For the RN case, however, there is again isospectrality of the axial and polar EM-led modes,
and there is also isospectrality of the axial and polar grav-led modes.
Thus, basically only three types of modes are left,
which all show a very similar pattern.
It also resembles the pattern seen for $l=0$ and $l=1$.
We will therefore now focus the discussion on the more interesting 
and varied behavior of the modes of the EMs black holes. 

As seen in Fig.~\ref{fig:l2_scalar_R},
the scalar-led $l=2$ modes change monotonically on the cold EMs branch.
The frequency $\omega_R/M$ decreases, and the damping rate $|\omega_I/M|$
increases toward the bifurcation point $q_{\text{min}}$.
Along the hot EMs branch, the frequency $\omega_R/M$ rapidly reaches a minimum
and then increases, while the damping rate $|\omega_I/M|$ exhibits a sinusoidal
behavior. At the extremal EMs solution, both the frequency and the damping rate
are higher than that at the extremal RN solution.

The EM-led $l=2$ modes are exhibited in Fig~\ref{fig:l2_EMap_R}.
They show a pattern that is very similar to the pattern of the EM-led $l=1$ modes,
although the numerical values of the frequencies and damping rates differ, of course.
The grav-led $l=2$ modes are exhibited in Fig.~\ref{fig:l2_gravap_R}.
The frequencies $\omega_R/M$ on the axial and polar cold EMs branches
follow very closely the RN branch almost up to 
the bifurcation point $q_{\text{min}}$.
This also holds for the damping rate $|\omega_I/M|$ on the axial cold EMs branch.
Only the damping rate along the polar EMs branch starts to deviate from the
RN one somewhat earlier.
Along the hot EMs branch, the frequencies $\omega_R/M$ show opposite
behavior with the frequency increasing monotonically on the axial branch,
while decreasing monotonically along the polar branch.
The damping rate $|\omega_I/M|$ also increases monotonically along
the axial branch, whereas it exhibits a strong sinusoidal behavior
along the polar branch.

\subsection{Isospectrality}

\begin{figure}[t]
	\centering
	\includegraphics[width=0.35\linewidth,angle=-90]{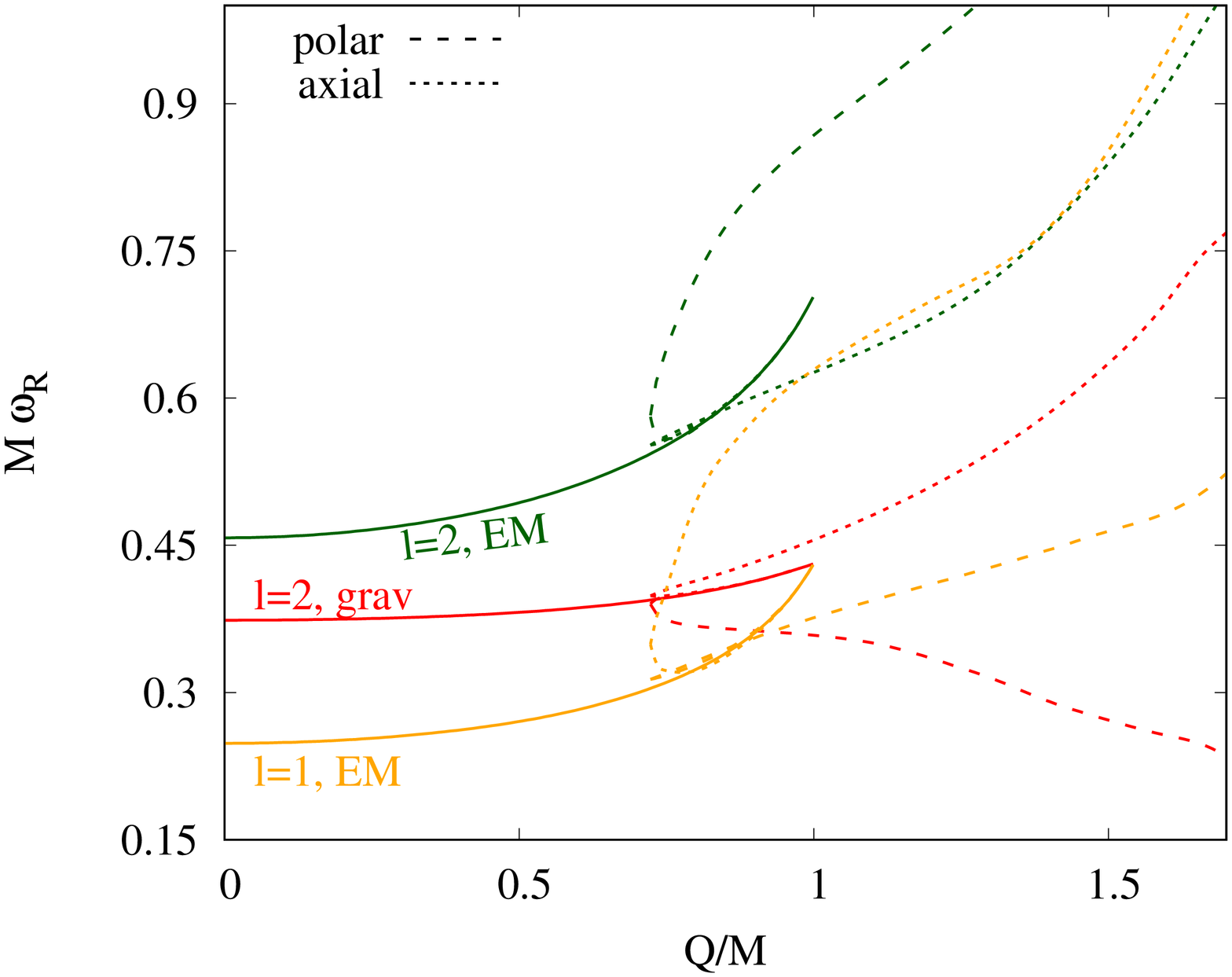}
      \includegraphics[width=0.35\linewidth,angle=-90]{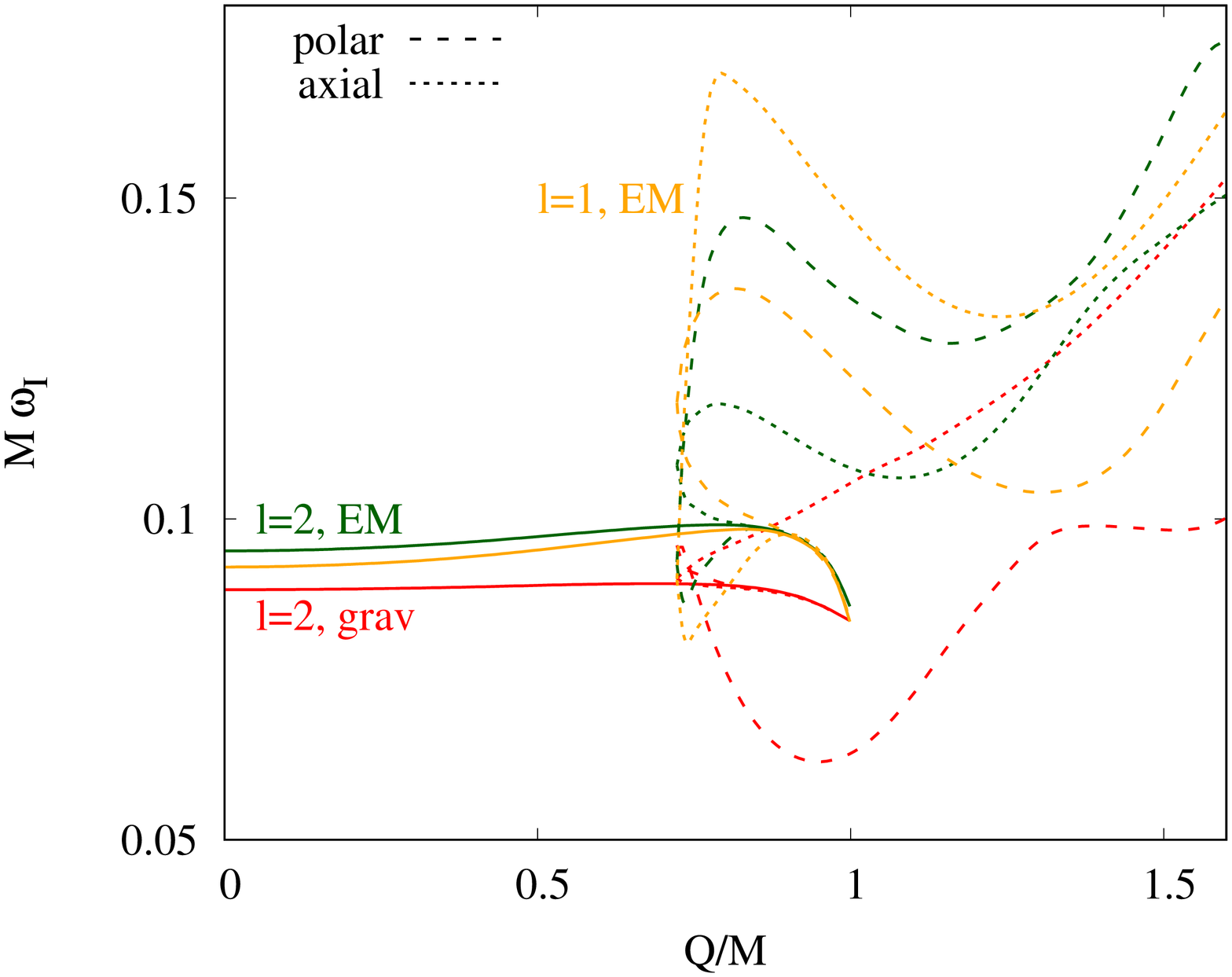}
	\caption{Isospectrality and its breaking for  
	axial (dotted) and polar (dashed) EM-led ($l=1$ yellow, $l=2$ green) and grav-led ($l=2$ red) perturbations:
	scaled frequency $\omega_R/M$ (a) and scaled damping rate $-\omega_I/M$ (b) 
       vs. reduced charge $q$
       for RN, and cold and hot scalarized black holes ($\alpha=200$).}
	\label{fig:iso1}
\end{figure}

As addressed in the above discussion, for RN black holes, 
isospectrality of axial and polar quasinormal modes holds. Here, axial and polar modes
coincide for EM-led modes as well as for grav-led modes, for any allowed
angular number $l$.
Moreover, in the extremal case, the EM-led modes with angular number $l$
agree closely with the grav-led modes with angular number $l+1$
\cite{Onozawa:1995vu,Andersson:1996xw}.
As seen above and once more highlighted in Fig.~\ref{fig:iso1},
this isospectrality survives to a large extent along the cold EMs branch,
since the EMs modes follow the RN modes closely over a large range of their existence.
In view of Fig.~\ref{fig:bh}, this may, however, not be too surprising,
since the cold EMs branch itself follows closely the RN branch
almost up to the bifurcation point $q_{\text{min}}$.

In contrast, the hot EMs branch reaches far beyond the RN branch
and thus also far beyond the cold EMs branch.
Its modes are therefore expected to show a generic behavior on their own.
In particular, this branch features a significant background scalar field.
The presence of a background scalar field, however, leads to 
a coupling of all the different perturbations, scalar, vector and gravitational
in the polar modes, whereas the axial modes remain free of
scalar perturbations.
Therefore it should not be surprising that isospectrality gets strongly
broken along the hot EMs branch, as illustrated in Fig.~\ref{fig:iso1}.

The breaking of isospectrality in the presence of a scalar field
has been observed before, of course, notably in Einstein-Maxwell-dilaton
black holes \cite{Blazquez-Salcedo:2019nwd,Ferrari:2000ep,Brito:2018hjh}, 
and Einstein-scalar-Gauss-Bonnet black holes
\cite{Blazquez-Salcedo:2020rhf,Blazquez-Salcedo:2020caw,Blazquez-Salcedo:2016enn,Blazquez-Salcedo:2017txk}.

\section{Conclusion}

EMs theories represent an interesting simple setting to study 
generic properties of spontaneous scalarization of black holes { and, more generically, the interplay between bald and hairy black holes.}
Although the {presence of scalar hair}
 is charge-induced,
many similarities with the more realistic curvature-induced
spontaneous scalarization {and scalar hairy black holes} have been observed,
lending weight to EMs studies beyond an intrinsic theoretical interest.

Here, we have investigated the linear mode stability of EMs
black hole solutions with a quadratic coupling function,
representing type IIB characteristics:
the presence of scalarized black black holes,
while the RN branch remains stable throughout.
In particular, the system features besides the bald RN branch
a cold EMs branch and a hot EMs branch.
The cold EMs branch exists 
in the interval $q_{\text{min}}(\alpha) \le q \le1$,
and the hot EMs branch 
in $q_{\text{min}}(\alpha) \le q \le q_{\text{max}}(\alpha)$.

The EMs black holes on the cold branch
has an unstable 
scalar-led mode, present for all $q_{\text{min}}(\alpha) \le q \le1$.
At $q_{\text{min}}(\alpha)$, this instability becomes a zero mode that is shared
with the hot branch. By continuity, the hot branch then
exhibits a stable scalar-led mode, starting at $q_{\text{min}}(\alpha)$.

In our mode analysis, we have calculated the lowest quasinormal modes
of all types of perturbations, scalar-led modes, 
axial and polar EM-led modes, and axial and polar grav-led modes
for all three (bald, cold, hot) branches.
For the stable modes on the cold EMs branch,
we have found close similarity with the respective RN modes,
almost up to the bifurcation point with the hot EMs branch $q_{\text{min}}(\alpha)$.
Since the cold EMs branch itself follows closely the RN branch,
this behavior could have been anticipated.
In contrast, the modes of the hot EMs branch show 
a wide variation and large deviations from the modes of RN branch.
But again, the hot EMs branch reaches far beyond the RN branch
(for sufficiently large $\alpha$).

RN black holes possess degenerate axial and polar modes
in the EM-led and grav-led sectors. As expected, this isospectrality 
gets broken in the presence of a non-trivial background scalar field,
since scalar perturbations contribute in the polar case but not in the axial case.
Not surprisingly, the breaking of isospectrality is very limited on the cold EMs branch,
but becomes very strong on the hot EMs branch, away from the bifurcation point.

The analysis here has focused on an illustrative value of the coupling $\alpha=200$ 
and for $\ell=0,1,2$, but the results concerning instability are generic for all values of $\alpha$; 
moreover, it is not to be expected that higher $l$ modes introduce more instabilities.
Thus our analysis has shown that there are no further unstable modes
apart from the unstable scalar-led $l=0$ mode of the cold EMs branch.
This implies that there are two linearly mode-stable branches in the system,
the bald RN branch and the hot EMs branch.
While both branches have large regions in parameter space,
where their black holes are the only existing black holes,
there is also an overlap region $q_{\text{min}}(\alpha) \le q \le1$.
Here, both branches coexist and both are mode-stable.

However, when these branches are considered from a
thermodynamical point of view, their reduced horizon areas differ
except at one critical point $q_{\text{th}}(\alpha)$, where the two curves cross,
as seen in Fig.~\ref{fig:bh}.
This might suggest that the RN black holes represent the physically preferred state
for $0 < q < q_{\text{th}}(\alpha)$,
whereas the hot EMs black holes represent the physically preferred state
for $q_{\text{th}}(\alpha) < q \le q_{\text{max}}(\alpha)$.
Here, dynamical calculations of the EMs evolution equations might
give further insight into the interesting question of which 
type of black hole will represent the end state of collapse in the overlap region.

\paragraph{Acknowledgments}
JLBS, SK and JK gratefully acknowledge support by the
DFG Research Training Group 1620  \textit{Models of Gravity}
and the COST Action CA16104 \textit{GWverse}.
JLBS would like to acknowledge support from
the DFG Project No. BL 1553. 
{AMP is supported by the FCT grant PD/BD/142842/2018. This work was supported by  the Center for Research and Development in Mathematics and Applications (CIDMA) through the Portuguese Foundation for Science and Technology (FCT - Funda\c c\~ao para a Ci\^encia e a Tecnologia), references UIDB/04106/2020 and UIDP/04106/2020, by national funds (OE), through FCT, I.P., in the scope of the framework contract foreseen in the numbers 4, 5 and 6
of the article 23, of the Decree-Law 57/2016, of August 29,
changed by Law 57/2017, of July 19 and by the projects PTDC/FIS-OUT/28407/2017 and CERN/FIS-PAR/0027/2019. This work has further been supported by  the  European  Union's  Horizon  2020  research  and  innovation  (RISE) programme H2020-MSCA-RISE-2017 Grant No.~FunFiCO-777740. We would like to acknowledge networking support by the COST Action GWverse CA16104.}

\appendix

\section{Perturbation equations}

In this appendix we present the explicit form of the perturbation equations
for the axial and polar case, and also the first terms in the asymptotic expansion of the perturbation functions.

\subsection{Axial perturbations}

By substituting the Ansatz (\ref{metric_axial_pert})
and (\ref{em_axial_pert}) into the field equations 
(\ref{Eins})-(\ref{Klein}), 
we obtain a set of differential equations for the axial perturbations:
\begin{eqnarray}
\label{eqh0}
 \partial_r h_0 = - i\omega h_1
 -4f(\phi_0)(\partial_r{a_{0}})W_2  +\frac{2}{r} h_{0}-{\frac {{i} }{2\omega\,{r}^{3}g  } h_{1}} \Big[ -\left( \partial_r\phi_{0}   \right) ^{2} g ^{2} {r}^{2}(r-2m) \nonumber \\-2rg^{2}l(l+1)
 +2rg\left( \partial_rg   \right)\left( r\partial_rm  + m - r \right) 
 -2{r}^{2}g\left( \partial_r^2 g   \right)(r-2m)   +4{r}^{2}{f(\phi_0)}g     \left( \partial_r{a_{0}}   \right) ^{2}(r-2m) \nonumber \\
  +\left( \partial_rg   \right) ^{2}  {r}^{2}(r-2m)+4\, \left( \partial_rm   \right) r g^{2}  +4g^2 (r-m) \Big] \ , \nonumber
\end{eqnarray}
\begin{eqnarray}
\partial_r h_1 =
{\frac {-i\omega r h_0 }{g  \left( r-2m \right) }} +h_1\frac{2g   \left( \partial_r m   \right) r-r\left( \partial_rg   \right) \left(r-2m\right) -2g  m}{2rg  \left( r-2m  \right) } \nonumber \ ,
\end{eqnarray}
\begin{eqnarray}
\partial_r^2{W_{2}}  =
{ \left(  \frac{4}{g}{f(\phi_0)}\left(\partial_r a_{0}\right)^{2}-\frac{{\omega}^{2}{r}}{g(r-2m)}+\frac{l(l+1)}{r(r-2m)} \right) {W_{2}} }  -\left[ \frac{\partial_r g}{2g} - \frac{\partial_rm}{r-2m} + \frac{\dot{f}(\phi_0)}{f(\phi_0)}\partial_r\phi_0 + \frac{m}{r(r-2m)}  \right] {\partial_r{W_{2}}} \nonumber \\  +\frac{i(\partial_r a_0) h_1}{2\omega r^3 g^2} \Big[ 4g^2(r-m) + 4rg^2\partial_rm - r^2g^2(r-2m)(\partial_r\phi_0)^2 - 2rg^2l(l+1)
\nonumber \\ 
+4gr^2f(\phi_0)(r-2m)(\partial_ra_0)^2 -2gr^2(r-2m)(\partial_r^2g)+r^2(r-2m)(\partial_rg)^2 + (2r^2g\partial_rm+2rg(m-r))\partial_rg\Big]
\nonumber \\ 
+h_0\left[\left(\frac{\partial_rg}{2g^2}+\frac{\partial_rm}{m(r-2m)}-\frac{\dot{f}(\phi_0)}{f(\phi_0)}\frac{\partial_r\phi_0}{g}+\frac{3m-2r}{gr(r-2m)}\right)\partial_ra_0-\frac{\partial_r^2a_0}{g}\right] \ . 
\end{eqnarray}

This system of coupled differential equations consists of
two first order differential equations for $h_0$ and $h_1$, plus a second order differential equation for $W_2$.

A perturbation with an outgoing wave behavior satisfying this system of equations has to behave like
\begin{eqnarray}
h_0 = r e^{i\omega R}  \left[A_g^+\left(-\omega+O(r^{-1})\right)+A_{F}^+\left(\frac{Q}{r^2}+O(r^{-3})\right)\right] \ ,  \nonumber \\
h_1 = r\omega e^{i\omega R} \left[A_g^+\left(1+O(r^{-1})\right)+A_{F}^+\left(\frac{-Q}{\omega r^2}+O(r^{-3})\right)\right] \ ,  \nonumber \\
W_2 = e^{i\omega R} \left[A_g^+\left(\frac{Q(l-1)(l+2)}{4\omega r^2}+O(r^{-3})\right)+A_{F}^+\left(1+O(r^{-1})\right)\right] \ .
\label{axout}
\end{eqnarray}

In addition, close to the horizon, a perturbation with an ingoing wave behavior has to satisfy
\begin{eqnarray}
h_0 &=& e^{-i\omega R}\left[A_g^-\left(1+O(r-r_H)\right) + A_{F}^-\left(O(r-r_H)\right)\right] \ ,  \nonumber \\
h_1 &=& \frac{\omega}{r-r_H}e^{-i\omega R}\left[A_g^-\left(\frac{r_H^{3/2}\sqrt{f(\phi_H)}}{\omega\sqrt{g_1\left(r_H^2f(\phi_H)-Q^2\right)}}+O(r-r_H)\right) + A_{F}^-\left(O(r-r_H)\right)\right] \ , \nonumber \\
W_2 &=& e^{-i\omega R}\left[A_g^-\left(O(r-r_H)\right)+A_{F}^-\left(1+O(r-r_H)\right)\right] \ . 
\label{axin}
\end{eqnarray}

The asymptotic expansion is determined by two independent amplitudes, one related to
the space-time perturbation $A_g^{\pm}$ 
and another related to the electromagnetic perturbation $A_{F}^{\pm}$.

\subsection{Polar case}

By substituting the Ansatz for the polar perturbations
(\ref{s_polar})-(\ref{metric_polar_pert})
into the field equations (\ref{Eins})-(\ref{Klein}), we obtain the following set of equations
\begin{eqnarray}
L+N=0 \ , \nonumber
\end{eqnarray}

\begin{eqnarray}
\partial_rH_1  =4{f(\phi_0)} \left( \partial_ra_0\right) (W_1-\partial_rV_1)  +\frac{\left(\partial_r\delta		  \right)r(r-2m)+2\left(\partial_rm   \right)r-2 m
}{r \left( r-2 m  \right)} H_1 
-\frac{2 i\omega r^{2}}{r \left( r-2 m  \right)}(T+L) \ , \nonumber
\end{eqnarray}

\begin{eqnarray}
\partial_r(N + T)  = {\frac {{r}^{3} \left( \partial_r \phi_0   \right) ^{2}(r-2m)f(\phi_0)-4{Q}^{2}  }{8	\left( r-2m   \right) {r}^{2}f(\phi_0)} (L-N)} 
+\frac{r-3m}{r(r-2m)}N
\nonumber \\
+\frac{r-m}{r(r-2m)}L
- \frac{2e^{2\delta}}{r(r-2m)}\left(Q(i\omega V_{1}+a_{1})-\frac{i}{4}\omega r^{2}H_{1}\right)+\frac{1}{2}\phi_1  (\partial_r \phi_0) \ , \nonumber
\end{eqnarray}

\begin{eqnarray}
\partial_rN =-\frac{r}{4} \left( \partial_r \phi_0   \right) (\partial_r\phi_{1}) 
- \frac{1}{r-2m} \Big[ \Big( -\left( \partial_r\delta   \right)r(r-2m)
- \left( \partial_rm \right) r-m  +r \Big) (\partial_rT)   -{e^{2\delta}f(\phi_0){r}^{2}\left(\partial_ra_0   \right)(\partial_ra_1)}
 \nonumber \\
-{i e^{2\delta}\omega rH_1}-L-\frac {1}{2} \Big(2{e^{2\delta}f(\phi_0)}{r}^{2} \left(\partial_ra_0\right)^{2}+l(l+1) \Big) N
-{\frac {1}{2} {e^{2\delta}\dot{f}(\phi_0)\left(\partial_ra_0\right)^{2}{r}^{2}\phi_1}} 
\nonumber \\
+{\frac{2{r}^{3}e^{2\delta}{\omega}^{2}-l(l+1)(r-2m)+2(r-2m)}{2\left(r-2m\right)}T  }-i \left( \partial_ra_0   \right) e^{2\delta}f(\phi_0)		\omega\,{r}^{2}W_1\Big] \ , \nonumber
\end{eqnarray}

\begin{eqnarray}
\partial_rT  ={\frac {i}{8\omega r^{2}} \left( 4{e^{2\delta}f(\phi_0)}{r}^{2} \left( \partial_r a_0 \right)^{2}+r\left( \partial_r \phi_0 \right) ^{2}(r-2m)+2{l}(l+1)-8(\partial_rm) \right) H_1 }
\nonumber \\
+{\frac {L}{r}}+\frac{1}{4}\phi_1  (\partial_r \phi_0)-{\frac{1}{r\left( r-2m\right)}\Big[r\left(\partial_r\delta   \right) (r-2m)+r((\partial_rm) +1) -3m\Big] T} \ , \nonumber
\end{eqnarray}

\begin{eqnarray}
\partial_r^2T  ={\frac {1}{r}}(\partial_rL)+ \frac{1}{4}\, \left( \partial_r \phi_0   \right) (\partial_r\phi_1)+\frac{1}{4r(r-2m)}\Big[ 4\left( 		r\left( \partial_rm   \right) +5\,m		 -3\,r \right) (\partial_rT) + {4 \left( \partial_ra_0   \right)   {e^			{2\delta}f(\phi_0)  }r^2(\partial_ra_1)  } \nonumber \\
+2{r^{2}  {e^{2\delta}}	\dot{f}(\phi_0)\left(\partial_ra_0   \right)^{2}\phi_1 }+{ \left( 4\,{			e^{2\delta}f(\phi_0)}{r}^{2} \left( \partial_ra_0   \right) ^{2}+r \left( \partial_r \phi_0   \right) ^{2}(r-2m)+2l(l+1)-4\,(2(\partial_rm)  		-1) \right) L  }
\nonumber \\
-{ \left(r \left( \partial_r		 \phi_0   \right) ^{2}(r-2m)  -8		\,\partial_rm   \right) N }+2\,{		\left( l+2 \right)  \left( l-1 \right) T  }+{4i \left( \partial_ra_0   \right)   {e^{2\delta}f(\phi_{0})				 } {\omega\,r^{2}W_1}}\Big] \ , \nonumber
\end{eqnarray}

\begin{eqnarray}
\partial_r^2(T+N)=
\frac{i\omega r e^{2\delta}}{r-2m}(\partial_rH_1)
-\frac{2-3r(\partial_r\delta)-4(\partial_rm)+6m(\partial_r\delta)}{r-2m}(\partial_rN) 
+ \frac{1}{2}(\partial_r\phi_0)(\partial_r\phi_1)
\nonumber \\
+ \frac{r(r-2m)(\partial_r\delta)+2m-2r+2r(\partial_rm)}{r-2m}(\partial_rT)
- \frac{2re^{2\delta}f(\phi_0)}{r-2m}(\partial_ra_0)(\partial_ra_1) 
- \frac{i\omega e^{2\delta}(m-r+r\partial_rm)}{(r-2m)^2}H_1
\nonumber \\
-\frac{2ri\omega e^{2\delta}f(\phi_0)(\partial_ra_0)}{r-2m}W_1
-\frac{r e^{2\delta}\dot{f}(\phi_0)(\partial_ra_0)^2}{r-2m}\phi_1
+\frac{e^{2\delta}\omega^2 r^2}{(r-2m)^2}(N-T)+\frac{2re^{2\delta}f(\phi_0)(\partial_ra_0)^2}{r-2m}T 
\nonumber \\
+\left[
2\partial_r^2\delta+\frac{2}{r-2m}\partial_r^2m-\frac{1}{2}(\partial_r\phi_0)^2-2(\partial_r\delta)^2+\frac{2(\partial_r\delta)(m+r-3r(\partial_rm))}{r(r-2m)}
\right](N+T) \ , \nonumber
\end{eqnarray}

\begin{eqnarray}
\partial_rV_1  =\frac {i{e^{2\delta}}\omega{r}^{3}}{l(l+1)(r-2m)}  \left[(\partial_ra_1)  +\left( \partial_ra_0   \right)\left(L+N-2T+\frac{\dot{f}(\phi_0)}{f(\phi_0)}\phi_1\right)\right] + \Big[1-\frac{\omega^2r^3e^{2\delta}}{l(l+1)(r-2m)}\Big]W_1 \ , \nonumber
\end{eqnarray}
\begin{eqnarray}
\partial_r^2V_1  = 
-{\frac{{r}^{2} e^{2\delta}{\omega}}{(r-2m)^{2}} \Big[\omega V_1-ia_1\Big]} 
+\partial_rW_1
\nonumber \\ 
-\frac {1}{r(r-2m)}  \left[2(m-r\partial_r m)+r(r-2m)\left(\frac{\dot{f}(\phi_0)}{f(\phi_0)}(\partial_r \phi_0)-(\partial_r \delta)\right)  \right] [\partial_rV_1-W_1] \ , \nonumber
\end{eqnarray}

\begin{eqnarray}
\partial_r^2a_1  =  -i\omega(\partial_rW_1)  - \left(\partial_ra_0\right) \left[\partial_r\left(L+N-2T+\frac{\dot{f}(\phi_0)}{f(\phi_0)}\phi_1\right)\right]  
 \nonumber \\
 +{\frac{l(l+1)}{r \left( r-2m \right) }}(i\omega V_1+a_1)-\left(\frac{\dot{f}(\phi_0)}{f(\phi_0)}\left( \partial_r \phi_0\right) +(\partial_r\delta)+\frac{2}{r} \right)\big[(\partial_r a_1)+i\omega W_1\big]
\nonumber \\
-{\left[\left( \partial_r a_0   \right)  \left( \partial_r \phi_0   \right) \frac{\dot{f}(\phi_0)}{f(\phi_0)}+ \left( \partial_r \delta   \right) (\partial_r a_0)  +(\partial_r^2a_0)  +\frac {2}{r}(\partial_ra_0)\right] \left[\frac{\dot{f}(\phi_0)}{f(\phi_0)}\,\phi_1+2(L+N)\right] } \ , \nonumber
\end{eqnarray}

\begin{eqnarray}
\partial_r^2\phi_1  =\left( \partial_r \phi_0   \right) \Big[\partial_r(N+2T-L)\Big]
-\frac{ir\omega e^{2\delta}\partial_r\phi_0}{r-2m}H_1-\frac{4r e^{2\delta}\dot{f}(\phi_0)}{r-2m}N-\frac{4ri\omega  e^{2\delta}\dot{f}(\phi_0)}{r-2m}W_1 \nonumber \\
-\frac{4r e^{2\delta}\dot{f}(\phi_0) (\partial_ra_0)}{r-2m} (\partial_ra_1)+\frac{2r(\partial_rm)+2m-2r+r(r-2m)(\partial_r\delta)}{r(r-2m)}(\partial_r\phi_1) \nonumber \\
+\left[2\partial_r\phi_0\partial_r\delta-2\partial_r^2\phi_0-\frac{4r e^{2\delta}\dot{f}(\phi_0)(\partial_ra_0)^2}{r-2m}+\frac{4(\partial_r\phi_0)(m-r+r\partial_rm)}{r(r-2m)}\right]L \nonumber \\
+\left[\frac{l(l+1)}{r(r-2m)}-\frac{r^2\omega^2e^{2\delta}}{(r-2m)^2}-\left(\frac{\dot{f}(\phi_0)}{f(\phi_0)}\right)^2\frac{2re^{2\delta}f(\phi_0)(\partial_ra_0)^2}{r-2m}\right]\phi_1 \ .
\end{eqnarray}

An useful redefinition of the electromagnetic perturbations is 
\begin{eqnarray}
\label{redef}
F_0(r) &=& -i\omega W_1(r) - \frac{dW_1(r)}{dr} \ , \\
F_1(r) &=& -i\omega V_1(r) - a_1(r) \ , \\
F_2(r) &=& -W_1(r) + \frac{dV_1(r)}{dr} \ ,
\end{eqnarray}
which allows to simplify the system of equations in a way that is more convenient for numerical calculations. After some algebra, the minimal set of differential equations, or
Master equations, can be written in vectorial form by defining a $6\times 6$ matrix (see equation (\ref{polar_master_Eq})). 

Polar perturbations with an outgoing wave behavior at infinity satisfying the previous equations behave like
\begin{eqnarray}
&&H_1 = r e^{i\omega R}  \left[A_g^+\left(-2+O(r^{-1})\right)+A_{F}^+\left(\frac{2iQ}{\omega r^2}+O(r^{-3})\right)+A_{\phi}^+\left(\frac{-i}{4\omega^2 r^4}\left(\omega\left.\frac{\dot{f}}{f}\right|_\infty Q^2-iQ_S\right)+O(r^{-5})\right)\right] \ ,  \nonumber \\
&&T = e^{i\omega R}  \left[A_g^+\left(1+O(r^{-1})\right)+A_{F}^+\left(\frac{-iQ}{\omega r^2}+O(r^{-3})\right)+A_{\phi}^+\left(\frac{iQ_S}{4\omega r^3}+O(r^{-4})\right)\right] \ ,  \nonumber \\
&&F_0 = \frac{1}{r^2} e^{i\omega R}  \left[A_g^+\left(-2Q+O(r^{-1})\right)+A_{F}^+\left(\frac{-il(l+1)}{\omega}+O(r^{-1})\right)+A_{\phi}^+\left(\left.\frac{\dot{f}}{f}\right|_\infty\frac{Q}{r}+O(r^{-2})\right)\right] \ ,  \nonumber \\
&&F_1 = e^{i\omega R}  \left[A_g^+\left(\frac{Q}{r}+O(r^{-2})\right)+A_{F}^+\left(1+O(r^{-1})\right)+A_{\phi}^+\left(-\left.\frac{\dot{f}}{f}\right|_\infty \frac{Q}{4r^2}+O(r^{-3})\right)\right] \ ,  \\
&&\phi_1 = \frac{1}{r}e^{i\omega R}  \left[A_g^+\left(\frac{2Q_SM}{r}-\left.\frac{\dot{f}}{f}\right|_\infty\frac{2 Q^2}{r}+O(r^{-1})\right)+A_{F}^+\left( \frac{Q l(l+1)}{\omega^2r^2}-\frac{i QQ_S}{\omega r^2}+O(r^{-3})\right)+A_{\phi}^+\left(1+O(r^{-1})\right)\right] \ . \nonumber
\label{polout}
\end{eqnarray}
Note that, since the background solutions we have considered satisfy $\phi(\infty)=0$, and the theory coupling is such that $\left.\frac{\dot{f}}{f}\right|_\infty=0$, some of the terms in this expansion vanish.

On the other hand, polar perturbations with an ingoing wave behavior at the horizon satisfy
\begin{eqnarray}
H_1 = \frac{1}{r-r_H} e^{-i\omega R}  \left[A_g^-\left(
\frac{2ir_H^2(2ie^{\delta_H}f(\phi_H)\omega r_H^3+f(\phi_H) r_H^2-Q^2)}{(2r_H e^{\delta_H}\omega+il(l+1))(f(\phi_H)r_H^2-Q^2)} + O(r-r_H)\right)
\right. \nonumber \\ \left.
+A_{F}^-\left(\frac{	4e^{\delta_H}f(\phi_H)Q r_H^3}{(r_H^2f(\phi_H)-Q^2)l(l+1)}(r-r_H)+O((r-r_H)^2)\right)
\right. \nonumber \\ \left.
+A_{\phi}^-\left(\frac{\dot{f}(\phi_H)}{f(\phi_H)}\frac{-4 r_H Q^2 r_H^3}{(r_H^2f(\phi_H)-Q^2)l(l+1)}(r-r_H)+O((r-r_H)^2)\right)\right] \ ,  \nonumber \\
T = e^{-i\omega R}  \left[A_g^-\left(
1 + O(r-r_H)\right)
\right. \ \ \ \ \ \ \ \ \ \ \ \ \ \ \ \ \ \ \ \ \ \ \ \ \ \ \ \ \ \ \ \ \ \ \ \ \ \ \  \ \ \ \ \ \ \ \ \ \ \ \ \ \ \ \ \ \ \ \nonumber \\ \left.
+A_{F}^-\left(\frac{-2e^{\delta_H}f(\phi_H)Q r_H(2r_H e^{\delta_H}\omega+il(l+1))}{(2r_H^3e^{\delta_H}f(\phi_H)\omega+if(\phi_H)r_H^2-iQ^2)l(l+1)}(r-r_H)+O((r-r_H)^2)\right)
\right. \nonumber \\ \left.
+A_{\phi}^-\left(\frac{\dot{f}(\phi_H)}{f(\phi_H)}\frac{- Q^2(2r_H e^{\delta_H}\omega+il(l+1))}{(2r_H^3e^{\delta_H}f(\phi_H)\omega+if(\phi_H)r_H^2-iQ^2)l(l+1)}(r-r_H)+O((r-r_H)^2)\right)\right] \ ,  \nonumber \\
F_0 = e^{-i\omega R}  \left[A_g^-\left(O(r-rH)\right)+A_{F}^-\left(1 + O(r-r_H)\right)+A_{\phi}^-\left(O(r-rH)\right)\right] \ ,  \nonumber \\
F_1 = e^{-i\omega R}  \left[A_g^-\left(\frac{-2i\omega  }{f(\phi_H)l(l+1)}+O(r-r_H))\right)+A_{F}^-\left(\frac{-ie^{\delta_H}\omega r_H^2}{l(l+1)}+O(r-rH)\right)
\right. \nonumber \\ \left.
+A_{\phi}^-\left(\frac{\dot{f}(\phi_H)}{f(\phi_H)^2}\frac{i\omega Q}{l(l+1)}+O(r-rH)\right)\right] \ ,  \nonumber \\
\phi_1 = e^{-i\omega R}  \left[A_g^-\left(O(r-rH)\right)+A_{F}^-\left(O(r-rH)\right)+A_{\phi}^-\left(1 + O(r-r_H)\right)\right] \ . 
\label{polin}
\end{eqnarray}

The asymptotic expansion of the polar perturbations is characterized by three undetermined amplitudes: $A_g^{\pm}$ (space-time perturbation amplitude), $A_{F}^{\pm}$ (electromagnetic perturbation amplitude)
and $A_{\phi}^{\pm}$ (scalar perturbation amplitude).

\bibliographystyle{ieeetr}
\bibliography{notes}

\end{document}